\documentclass[final,5p,times,twocolumn]{elsarticle}
\usepackage{url}
\usepackage{bbm}
\usepackage{amsmath,amsfonts} % Math packages
\usepackage{float}
\usepackage{graphicx}
\usepackage{algorithm}
\usepackage{algpseudocode}
\usepackage{siunitx}
\usepackage[labelfont=bf]{caption}
\usepackage{subcaption}
\usepackage{tikz}
\usepackage{pgfplots}
\usepackage{epstopdf}
\usepackage{multirow}
\usepackage[font=scriptsize]{caption}
\usepackage{soul}
\usepackage{color, colortbl}
\usepackage{array}
\usepackage[T1]{fontenc}
\usepackage[utf8]{inputenc}
\usepackage{mathptmx}
\usepackage{listings}
\usepackage{adjustbox}
\usepackage{makecell}
\usepackage{tcolorbox}

\usepackage{enumitem}
\usepackage{hyperref}
\usepackage{pifont}% http://ctan.org/pkg/pifont

\def\BibTeX{{\rm B\kern-.05em{\sc i\kern-.025em b}\kern-.08em
    T\kern-.1667em\lower.7ex\hbox{E}\kern-.125emX}}

\begin{document}

\begin{frontmatter}

\title{Evaluating The Explainability of Deep Learning-based Network Intrusion Detection Systems}

\author{Ayush Kumar, Vrizlynn L.L. Thing} %% Author name

%% Author affiliation
\affiliation{organization={Cyber Security Strategic Technology Centre, ST Engineering},%Department and Organization
            addressline={600 W Camp Rd}, 
            city={Singapore},
            postcode={797654}, 
            state={Singapore},
            country={Singapore}}

\begin{abstract}
State-of-the-art deep learning (DL)-based network intrusion detection systems (NIDSs) offer limited ``explainability''. For example, how do they make their decisions? Do they suffer from hidden correlations? Prior works have applied eXplainable AI (XAI) techniques to ML-based security systems such as conventional ML classifiers trained on public network intrusion datasets, Android malware detection and malicious PDF file detection. However, those works have not evaluated XAI methods on state-of-the-art DL-based NIDSs and do not use latest XAI tools. In this work, we analyze state-of-the-art DL-based NIDS models using conventional as well as recently proposed XAI techniques through extensive experiments with different attack datasets. Furthermore, we introduce a criteria to evaluate the level of agreement between global- and local-level explanations generated for an NIDS. Using this criteria in addition to other security-focused criteria, we compare the explanations generated across XAI methods. The results show that: (1) the decisions of some DL-based NIDS models can be better explained than other models, (2) XAI explanations generated using different tools are in conflict for most of the NIDS models considered in this work and (3) there are significant differences between XAI methods in terms of some security-focused criteria. Based on our results, we make recommendations on how to achieve a balance between explainability and model detection performance.
\end{abstract}

%\begin{IEEEImpStatement}
%Network Intrusion Detection Systems which use deep learning models achieve high detection performance and accuracy while avoiding dependence on fixed signatures extracted from attack artifacts. Despite that, recent survey-based research has shown that there is a noticeable hesitance among network security experts and practitioners when it comes to deploying DL-based NIDSs in real-world production environments due to their black-box nature, i.e., how and why the underlying models make their decisions. Our work paves the way for extracting human-understandable explanations for the decision-making behaviour of DL-based NIDSs, thus helping security practitioners trust them and confidently deploy them. The explanations could also enable security solution providers who use DL-based NIDS products as part of their solutions to satisfy clients who wish to know the reasons behind ``bad'' incidents such as missed detection of an attack/malicious software. 
%\end{IEEEImpStatement} 


\end{frontmatter}

\section{Introduction}
\label{intro}
%\subsection{IoT/IoT security}
%The Internet-of-Things (IoT) \cite{iotsurvey} constitutes a vast network of sensor-equipped devices, often characterized by low power and modest processing capabilities. These devices seamlessly exchange data among themselves or with central systems such as gateways or cloud servers using a mix of wired (such as Ethernet) and wireless technologies (such as RFID, Zigbee, WiFi, Bluetooth, and 3G/4G). The applications of IoT devices span a wide spectrum: from wearables like fitness trackers and smartwatches, to home automation systems managing lighting, thermostats, cameras, and door locks, all the way to industrial automation for tasks like manufacturing equipment management, process control, and ensuring plant safety. By 2032, experts estimate that a staggering 34.4 billion  IoT devices will be deployed globally \cite{txformaforecast}. Unfortunately, cyber attackers have set their sights on IoT devices, leveraging malicious software (commonly known as malware) \cite{iotsecsurvey1,iotsecsurvey2}. Their motivations stem from several factors: first, the prevalence of legacy devices still connected to the Internet without adequate security updates; second, the tendency to prioritize other aspects over security during the development cycle; and third, the widespread use of weak login credentials. The sheer volume of deployed devices ranging from cameras to sensors, poses a formidable challenge when it comes to implementing cost-effective, real-time anomaly detection at such a massive scale.

%\subsection{NIDS/DL-based NIDS} 
%from ENIDIFT paper
Securing networks against intrusion is paramount, and network intrusion detection systems (NIDS) play a pivotal role in achieving this. Traditionally, NIDSs have relied heavily on signature-based methods \cite{bro, snort}- essentially matching attack patterns to known signatures. However, these approaches demand significant manual effort from experts to create and maintain those signatures. Moreover, they fall short when it comes to identifying novel, previously unseen attacks. With the introduction of deep learning (DL)-based NIDSs which leverage DL models to enhance existing defenses, impressive performance and accuracy has been achieved \cite{dd-nids, MLIDissues} and they have been found to perform better than conventional machine learning (ML) algorithms \cite{kitsune}.
%Unlike their signature-centric counterparts, DL-based NIDSs do not require experts to painstakingly prepare signatures in advance. 
In fact, unsupervised DL-based NIDSs can even raise alerts for hitherto unknown attacks. Thanks to these advantages and the rapid evolution of DL techniques, DL-based NIDSs have made significant strides in recent years.

%\subsection{Need for explaining IoT NIDS decisions}
%from TRUSTEE paper
Despite the promise of DL-based NIDSs, there is a noticeable hesitance among network security experts and practitioners when it comes to deploying them in real-world production environments \cite{dos-n-donts-ml}. The crux of the issue lies in the black-box nature of many of these proposed solutions. Essentially, their inner workings (how and why they make decisions) remain elusive, unlike simpler but less effective rule-based approaches that security professionals are accustomed to. Security analysts rely on DL-based NIDSs for critical decisions. However, without clear explanations for the decisions made by these underlying DL models, it becomes difficult for the analysts to trust them. Moreover, DL often operates as a black box within the NIDS. Consequently, hidden correlations stemming from unrelated artifacts can slip past undetected, leading to a flawed perception of the system’s capabilities. Explanation tools and techniques can unveil those hidden correlations, allowing practitioners to assess their impact on the NIDS’s overall effectiveness. As it stands, the current state-of-the-art DL-based NIDSs lack inherent explainability. Therefore, there is a pressing need to scrutinize them using these very explanation techniques. 

%\subsection{Brief description of our solution}
In this work, we analyze state-of-the-art DL-based NIDS models using eXplainable AI (XAI) techniques (e.g., TRUSTEE \cite{trustee}, SHAP \cite{shap}). In addition, we develop a criteria to evaluate the level of agreement between global- and local-level explanations generated for an NIDS. We compare the explanations generated across XAI methods using our criteria and other security-focused evaluation criteria. To the best of our knowledge, our work is the first one to generate explanations for state-of-the-art DL-based NIDSs other than Kitsune using XAI techniques. Though a few works \cite{kalakoti, patil, arreche, barnard, warnecke} have applied XAI techniques to ML-based security systems (conventional ML classifiers trained on public network intrusion datasets, Android malware detection, malicious PDF file detection) in the past, none of those systems corresponded to state-of-the-art DL-based NIDSs.

The main contributions of our work are as follows:
\begin{itemize}
	\item We analyze state-of-the-art DL-based NIDS models using XAI techniques. The application of XAI techniques is not trivial and we make several modifications to the NIDS implementations.  
	\item We propose a criteria to evaluate the level of agreement between global- and local-level explanations generated for an NIDS.
	\item We compare the explanations generated across XAI methods using our criteria and other security-focused evaluation criteria. This highlights the need (for a security practitioner) to select the best method for a particular NIDS as per the preferred criteria. 
	\item We propose an approach (for a security practitioner) to select the best DL-based NIDS for a specific use-case by achieving the desired balance between detection performance and explainability. 
%	\item Using the above explanations and metrics, we show that: 
%	\begin{itemize}
%		\item Some DL-based NIDS models can be better explained (in terms of interpretable decision trees) than other models.
%		\item XAI explanations generated using different tools (e.g., TRUSTEE, SHAP) are in conflict with each other for most of the NIDS models considered.
%		\item Some NIDS models are more vulnerable to inductive bias (in terms of learnt packet rates) than other models.
%	\end{itemize}
\end{itemize}
 
The rest of the paper is organized as follows. In Section \ref{background}, we present a background on the types of explanation methods for ML models, prominent black-box XAI methods, and an overview of the state-of-the-art DL-based NIDSs to which we apply explanation techniques in this work. In Section \ref{XAI-analysis}, we explain the issues with applying XAI methods to the DL-based NIDSs directly using their open-source implementations, our approach to addressing those issues and details of our investigation beyond direct application of XAI methods which covers the relation between explanations generated by different XAI methods. In Section \ref{results}, we evaluate the explanations generated for each of the NIDSs using XAI methods, compare the explanations across XAI methods, and in Section \ref{discuss}, we discuss insights gleaned from the main results obtained thus far. In Section \ref{literature}, we review published research on the need for explanation of ML systems targeted at cybersecurity, evaluation of explanation methods on DL-based security systems and explainable DL-based attack detection. Finally, we summarise the key results obtained and provide future research directions in Section \ref{conclusion}. 

\section{Background}
\label{background}

In this paper, since we delve into explaining the decisions of state-of-the-art DL-based NIDSs, it is essential to briefly explore the types of explanation methods present in academic literature, the NIDSs that we consider for explanation in this work and the various aspects of our investigation with XAI techniques on those NIDSs. 
  
\subsection{Explanation Methods- Types}
All explanation methods for ML/DL models can be broadly categorized into \textit{black-box} and \textit{white-box} explanations.

\textbf{Black-box Explanations} These methods function within a black-box context, assuming zero knowledge about the inner workings of the underlying ML model and its associated parameters. These methods prove invaluable when direct access to the ML model is not available, for instance, during remote audits of a learning-based application. These black-box methods approximate the underlying ML model’s prediction function $f_N$, allowing them to estimate how various dimensions of the input vector $x$ affect a given prediction. While they hold promise for explaining deep learning models, their effectiveness can be impeded by the black-box environment, sometimes leaving out crucial insights embedded in the ML model’s parameters and architecture.

\textbf{White-box Explanations} These methods work under the premise that we have full knowledge of all the parameters of the ML model being tested. Armed with this information, these methods bypass approximations and can directly compute explanations for the prediction function, $f_N$. The same ML model is often used to generate both predictions and explanations, assuming that the model is available on hand. This assumption holds true for stand-alone automated systems such as those that analyse binaries, detect malware or discover software vulnerabilties. However, it is worth noting that some white-box methods are customized for specific neural network architectures deployed in the fields of computer vision/speech processing/natural language processing and may not perform well for other architectures.

\subsection{Prominent Black-box XAI Tools}
\label{background-XAI}
DL-based NIDS products are expected to be used by end users/customers who do not have access to the ML model and its parameters underlying the NIDS. This is to maintain confidentiality of the NIDS implementation by the cybersecurity services provider. Therefore, we assume that the DL-based NIDSs being explained in this work are available to us as a black-box system only and we use XAI techniques designed to explain such black-box ML systems. In what follows, we present a brief overview of each black-box XAI tool that we have tested with as part of this work.

\textbf{Overview of TRUSTEE}: TRUSTEE \cite{trustee} is a state-of-the-art interpretability tool for ML/DL-based security solutions which generates interpretable decision trees (DT) for a given ML/DL model. It can be used to find out inductive biases in black-box models (e.g., shortcut learning, spurious correlations). While generating DT, TRUSTEE focuses on the following desirable properties: model-agnosticism (should be applicable to any given black-box model), model fidelity (how accurate the DT is with respect to the original ML model), model stability (the decision rules should correctly describe how the given black-box model makes a significant number of its decisions and should be largely insensitive to the particular data samples that TRUSTEE used in the process of selecting its final DT explanation) and model complexity (decision rules should be readily recognizable by domain experts). The important aspects of generated DT explanation are summarized in a trust report which helps end users determine whether the given black-box model suffers from the problem of underspecification and can be trusted.

In the TRUSTEE algorithm, there is an inner loop which is designed to generate different high fidelity DT explanations, one per iteration. It does so by applying a teacher-student dynamic derived from imitation learning that uses the blackbox model, $\pi^*$ as an oracle in conjunction with a carefully curated dataset $D’$ to guide the training of a surrogate ``white-box'' model in the form of a DT that imitates the black-box’s decisions. In contrast, the purpose of the outer loop is (i) to select from among the $N$ high fidelity DTs that have been generated in the process of executing the inner loop the DT with the highest fidelity, (ii) to transform this resulting DT into a high-fidelity and low-complexity DT by means of a post-processing step that consists of applying a purposefully developed pruning method called top-$k$ pruning which carefully scrutinizes only the top $k$ branches of an extracted high-fidelity DT, ranked by the number of input samples a branch classifies, and (iii) to consider all $S$ high-fidelity and low-complexity DTs that have been generated in the process of executing the outer loop and output the one that is the most stable in the sense of having the highest mean agreement among these $S$ DTs.

\textbf{Overview of LIME}: Ribeiro et al. \cite{lime} introduced LIME, one of the first black-box methods for explaining neural networks that is further extended by SHAP \cite{shap}. Both works are motivated by the need to allow users to be able to trust ML models as well as their predictions. LIME is designed to explain the predictions of any classifier or regressor in a faithful way, by approximating it locally with an interpretable model. 
%The main characteristics of LIME are:
%\begin{itemize}
%    \item Interpretable, i.e., it provides qualitative understanding of the relation between the input variables and the response.
%    \item Locally faithful, i.e., it corresponds to how the model behaves in the vicinity of the instance being predicted.
%    \item Model-agnostic, i.e., it treats the original ML model as a black box.
%\end{itemize}
LIME aims at approximating the decision function, $f$ of the model being explained by solving the following optimization problem:
\begin{equation}
\underset{g\in \mathcal{G}}{arg min} \ \mathcal{L}(f, g, \pi_x),
\end{equation}
where $\mathcal{L}(f, g, \pi_x)$ is the fidelity function which is a measure of how unfaithful the interpretable model $g$ is in approximating $f$ in the locality defined by $\pi_x$, while $\pi_x(z)$ is the proximity measure between an instance $z$ to $x$, so as to define the locality around $x$ and $x$ is the input to $f$.

In order to learn the local behavior of $f$ as the interpretable inputs vary, LIME approximates $\mathcal{L}(f, g, \pi_x)$ by drawing samples, weighted by $\pi_x$. The instances around $x$ are sampled by drawing nonzero elements of $x$ uniformly at random. Next, LIME creates a series of $l$ perturbations of $x$, denoted as $\tilde{x_1}$, $\dots$, $\tilde{x_l}$ by setting entries in the vector $x$ to $0$ randomly. The method then proceeds by predicting a label $f(\tilde{x_i}) = \tilde{y_i}$ for each $\tilde{x_i}$ of the $l$ perturbations. This strategy enables the method to approximate the local neighborhood of $f$ at the point $f(x)$. LIME approximates the decision boundary by a weighted linear regression model,
\begin{equation}
\mathcal{L}(f, g, \pi_x) = \sum_{i=1}^l \pi_x(\tilde{x_i})(f(\tilde{x_i})-g(\tilde{x_i}))^2 ,
\end{equation}
where $\mathcal{G}$ is the set of all linear functions and $\pi_x$ is a function indicating the difference between the input $x$ and a perturbation $\tilde{x}$. 

\textbf{Overview of SHAP}: SHAP \cite{shap} (SHapley Additive exPlanations) subsumes additive feature attribution methods such as LIME and uses the following forms for $\mathcal{L}(f, g, \pi_x)$ and $\pi_x$ under the SHAP kernel method, which is shown to recover Shapley Values \cite{shapley-val} when solving the regression:
\begin{eqnarray*}
\mathcal{L}(f, g, \pi_{x'}) = \sum_{z' \in Z} [h_x^{-1}(z')-g(z')]^2 \pi_{x'}(z'), \\
\pi_{x'}(z') = \frac{M-1}{(M \text{choose} |z'|)|z'|(M-|z'|)} 
\end{eqnarray*}
where $x'$ is the simplified input which maps to the original input $x$ through a mapping function $x = h_x(x')$, $M$ is the number of simplified input features, $z' \in {\{0,1\}}^M$, and $|z'|$ is the number of non-zero elements in $z'$. 
%The above solution satisfies the properties of local accuracy (the explanation model should at least match the output of $f$ for the simplified input $x'$ ), missingness (if the simplified inputs represent feature presence, features missing in the original input should have no impact) and consistency (if a model changes so that some simplified input’s contribution increases or stays the same regardless of the other inputs, that input’s attribution should not decrease).

Shapley values are a concept from game theory where the features act as players under the objective of finding a fair contribution of the features to the payout, in this case the prediction of the model. Different from TRUSTEE, Shapley values are used for local interpretation. In other words, it is used to interpret how a prediction result is reached for a specific data sample or subset of samples. Specifically, Shapley values can tell how each feature contributes to the predicted results. A positive Shapley value means that the feature’s value pushes the classification result toward being malicious, and a negative Shapley value does the contrary. 

\textbf{Overview of LEMNA}: LEMNA \cite{lemna} is a black-box method specifically designed for security applications. Given an input data instance $x$ and a classifier such as an RNN, this method aims to identify a small set of features that have key contributions to the classification of $x$. This is done by generating a local approximation of the target classifier’s decision boundary near $x$. It uses a mixture regression model (to approximate locally non-linear decision boundaries) enhanced by fused lasso (to handle correlated features) for approximation. The mixture regression model is a weighted sum of $K$ linear regression models:
\begin{equation}
f(x) = \sum_{j=1}^K \pi_j(\beta_j \cdot x + \epsilon_j).
\end{equation}
The parameter $K$ specifies the number of models, the random variables $\epsilon = (\epsilon_1, \dots , \epsilon_K)$ originate from a normal distribution $\epsilon_i \sim \mathcal{N}(0, \sigma)$ and $\pi = (\pi_1, \dots, \pi_K)$ holds the weights for each model. The variables $\beta_1, \dots , \beta_K$ are the regression coefficients and can be interpreted as $K$ linear approximations of the decision boundary near $f(x)$.

\subsection{State-of-the-art Deep Learning-based NIDSs}
\label{sota-dl-nids}
We now present a brief overview of state-of-the-art DL-based NIDSs that we have considered in this work. 
%All these NIDSs use packet-level features extracted from incoming raw packets except IoTA which uses automata constructed from traffic flows. 

\textbf{Kitsune} \cite{kitsune} proposes an online unsupervised anomaly detection system based on an ensemble of autoencoders which is lightweight in terms of memory footprint and meant to be deployed on network gateways and routers to detect attacks on the local network. It uses a damped incremental statistics (DIS) method to extract network traffic features which has the advantage of extracting features from dynamic network traffic at high speed. 

\textbf{HorusEye} \cite{horuseye} is a two-stage anomaly detection framework for IoT. In the first stage (Gulliver Tunnel), preliminary burst-level anomaly detection is implemented on the data plane using an isolation forest model converted into a set of white-listing rules. The suspicious traffic is then reported to the control plane for further investigation. To reduce the false positive rate, the control plane carries out the second stage (Magnifier), where more thorough anomaly detection is performed over the reported suspicious traffic using Kitsune DIS features and an asymmetric autoencoder with separable and dilated convolutions.

\textbf{ENIDrift} \cite{enidrift} addresses real-world issues in DL-based NIDS deployment such as concept drift (change in statistical properties of target variable over time), class-imbalanced training data and well-crafted ML attacks (training data contamination, adversarial attacks). It includes three main components: iP2V- an incremental feature extraction method based on Word2Vec, an ensemble of autoencoders, a sub-classifier generation module and an update module. 

\textbf{HELAD} \cite{helad} uses ensemble learning consisting of an unsupervised autoencoder as a base learner which learns the profile for normal network traffic and a supervised LSTM (Long-term short-term memory) which is trained using the autoencoder’s output to detect continuous attacks. The model is re-trained using a new concept of time slice re-training. It also uses dynamic thresholds and integrated learning parameters to keep the anomaly detection performance from degradation.

\textbf{Deep ResNIDS} \cite{resnids} uses a combination of three deep-neural network stages to detect novel attacks at the packet traffic-level: a malicious packet detector (supervised) which detects known malicious traffic, an anomaly detector (unsupervised) which distinguishes novel packets (out-of-distribution, zero-day and adversarial attacks) from truly benign packets and a novelty detector (supervised) which distinguishes adversarial manipulated packets from truly novel packets. To re-train the models in above three stages for addressing evolving attacks, a transfer learning approach is used wherein all but few of the trained layers' weights are frozen and an additional layer is appended to the model for training. 

%\textbf{IoTA} \cite{iot-a} is an intrusion detection system for IoT devices which uses fully packet-level models to profile IoT device traffic patterns. These models are based on automata constructed from raw device traffic traces whose states represent packets with similar context, representative features are packet length and direction and when grouped together based on connectedness, represent device traffic flows. At runtime, ongoing IoT device traffic is matched with the extracted normal traffic profiles towards intrusion detection.

The above state-of-the-art systems were selected since: (1) they were published in prominent avenues (NDSS, USENIX Security, ACSAC, Computers \& Security journal) and include works published as recently as 2024, (2) they provide a clear performance comparison with Kitsune (which is the most cited deep learning-based NIDS with an open-source implementation) and (3) they cover a diverse ground on the application of deep learning in network intrusion detection. Kitsune uses an ensemble of autoencoders while ENIDrift builds on top of that by including a sub-classifier generation and updation module to address concept drift in real-world networks. HorusEye's Magnifier switches the ensemble of autoencoders for a single asymmetric autoencoder to keep the solution lightweight and reduce the false-positive rate. HELAD uses a single autoencoder as well but combines it with an LSTM network. Deep ResNIDS employs three stages of deep-learning classifiers to detect novel attacks out of which the latter two stages (anomaly detector and novelty detector) use autoencoders.

\section{XAI Evaluation of Deep Learning-based NIDSs}
\label{XAI-analysis}
In this work, we analyze state-of-the-art DL-based NIDSs presented in previous section using XAI techniques. However, XAI techniques such as TRUSTEE and kernel SHAP can not be applied directly to the NIDSs under consideration. This is because XAI methods have certain requirements at the API level which are not fulfilled by the current implementations of the NIDSs. We had to modify the NIDS implementations to make them compatible with XAI methods. Still, there were a few XAI methods whose requirements were found to be incompatible with the NIDS implementations unless we made major changes to the source code. 
%Additionally, we did not limit ourselves to a direct application of XAI methods to the NIDSs but we also investigated: (1) the relation between the explanations generated for a given NIDS by different XAI methods, and (2) the presence of inductive biases in the NIDS implementations.

\subsection{XAI-compatible NIDS Implementation}
\label{XAI-nids-impl}
The TRUSTEE trust report generating API has two requirements: (1) a trained ML/DL model with a \textit{predict()} function taking training/test dataset as the only input and outputting predicted labels, and (2) training/test datasets and corresponding predicted labels should be in \textit{numpy} array/\textit{pandas} dataframe format. The kernel SHAP API shares the requirements of TRUSTEE API. Except Kitsune, other NIDS source codes do not provide trained ML model or training/test data in a format which can be directly input to TRUSTEE API or kernel SHAP API. 

For example, in the case of ENIDrift source code, the input dataset and predicted labels were not in \textit{numpy} array/\textit{pandas} dataframe format while the \textit{predict()} function was not designed for an input dataset in the form of a 2-D matrix with rows as samples and columns as feature values. Further, ENIDrift does not extract explicit features from incoming packets but instead embeds packets to vectors which may not be meaningful for model explainability analysis. In the case of HorusEye source code, the main script was in the form of a command with argument parsing, all library functions were defined in the same main script, there was no \textit{predict()} function and input dataset and predicted labels were not in \textit{numpy} array/\textit{pandas} dataframe format. In the case of HELAD source code, there was no \textit{predict()} function implemented and the NIDS model returned anomaly scores instead of class labels. Further, the way anomaly scores were calculated was not compatible with TRUSTEE API implementation.

While running LIME analysis of above NIDSs, we received an error that LIME does not support ML/DL models which do not have $predict\_proba()$ function implemented. Since all the state-of-the-art DL-based NIDSs considered in this paper do not implement $predict\_proba()$ function, we are forced to leave out LIME analysis of the NIDSs from this paper. Here, $predict\_proba()$ function returns the predicted probabilities of the input features belonging to each category. This method, instead of returning a discrete class, returns the probabilities associated with each class. This is useful when we not only want to know the predicted category of the input features, but also the model's confidence in its prediction. Since SHAP is a generalized version of LIME, excluding LIME from our analysis would not affect our results substantially.

While running LEMNA analysis of above NIDSs, we found that the current implementation (last changed 5 years ago) only supports basic MLP, CNN and RNN models. Custom ML/DL models such as the NIDS models which are under consideration in this work are not supported. Further, LEMNA does not output feature-based model explanations for processing by cybersecurity experts such as those offered by TRUSTEE and SHAP. Though LEMNA improves upon LIME by using non-linear local approximation and handling feature dependencies, it is still inferior to SHAP due to the latter's high mathematical rigor and strong guarantees such as consistency and local accuracy. Thus, excluding LEMNA from our analysis would not affect our results substantially. 

\subsection{Evaluation Criteria}
\label{trustee-shap-rel}
To evaluate the performance of various explanation methods on DL-based NIDSs, we propose a new criteria called \textit{TRUSTEE-SHAP Agreement Score}. Further, we use the above criteria as well as security-related criteria introduced in \cite{sok-dl-sec} (\textit{Stability} and \textit{Adversarial Robustness}) to compare the explanation methods. These criteria should come in handy for security practitioners while selecting the best explanation method suited to their use-case.

\subsubsection{\textbf{TRUSTEE-SHAP Agreement Score}} 
To establish the relationship between TRUSTEE global explanations and the SHAP local explanations of the black-box model underlying a given NIDS, we can follow two approaches:
\begin{itemize} 
	\item \textit{Na\"ive approach}- We can directly compare the kernel SHAP features with highest SHAP values against the top-k features identified by TRUSTEE report. However, this comparison would not be meaningful as the SHAP value for each feature is calculated for a subset of data samples and reflects the importance of that feature in the decision made by the black-box model for a majority of the data samples in that subset. On the other hand, TRUSTEE acts on the complete input dataset and identifies the top features contributing the most to forming the decision boundary separating all benign and malicious data samples which are part of the input dataset.
	\item \textit{Systematic approach}- Since XAI methods such as SHAP are more suited towards analyzing a subset of data samples \cite{trustee}, they can be applied to the subset of samples corresponding to a path in a TRUSTEE DT starting at the root node and ending at one of the leaf nodes. We can compare the explanation thus generated with the one provided by TRUSTEE.
\end{itemize}

Under the systematic approach, for a given subset of data samples $D$, if $T$ is the set of features used in the selected root node-leaf node DT path decisions and $S$ is the set of features identified by kernel SHAP that contribute most to the classification decision for majority of the samples in $D$, then $T \subset S$ should hold true for the TRUSTEE DT explanation to agree with the SHAP explanation of the black-box model at the subset level. To form the set of features $S$, the mean absolute value of kernel SHAP values for multiple data samples are calculated for every feature, and subsequently the features are ranked according to their aggregate SHAP values to find the top features. For a black-box NIDS model, the local explanation (SHAP) may agree with the global explanation (TRUSTEE) depending on the data subset on which they are executed. However, not all local explanations may agree with the global explanation. 

To evaluate the level of agreement between TRUSTEE and SHAP for an NIDS trained on a given dataset, we propose the following approach: 
\begin{enumerate}
    \item Sample different subsets from the original evaluation dataset by selecting different root node-leaf node paths in the TRUSTEE DT. 
	\item Use each sampled data subset for executing kernel SHAP on the trained NIDS. 
	\item We propose the \textit{TRUSTEE-SHAP agreement score} on a data subset $D$ which can be calculated as: 
	
    \begin{equation}
    	\alpha_{TRUSTEE-SHAP}^D = \frac{n(T \subset S)}{n(T)},
    \end{equation}
    where $n(T \subset S)$ refers to the number of TRUSTEE DT features which are a subset of SHAP features and $n(T)$ is the total number of TRUSTEE DT features corresponding to the selected data subset $D$.
\end{enumerate}

We define the \textit{m-point average TRUSTEE-SHAP agreement score} for an NIDS as the average of TRUSTEE-SHAP agreement scores for $m$ different data subsets. It can be expressed as:

\begin{equation}
    A_{TRUSTEE-SHAP}^m = \frac{1}{m} \sum_{m} \alpha_{TRUSTEE-SHAP}^{D_m}
\end{equation}

%\subsubsection{\textbf{Multi-class Differentiability}}:
%\subsubsection{\textbf{Semantic Expressibility}}: 

\subsubsection{\textbf{Adversarial Robustness}} 
Multiple works \cite{ghorbani, dombrowski, alvarez} have shown that small input perturbations can lead to significant changes in explanations while the classification output remains same. Another class of adversarial attacks against LIME and SHAP \cite{slack, lemerrer} replaces a black-box model (with biased predictions) with a surrogate model to hide the bias (e.g., from external auditors). A user querying the model service for the explanation of a decision by the black-box model is instead provided the explanations of the surrogate model. Though defensive strategies to increase the robustness of explanation techniques have been proposed in recent works, there is still a lack of formal analysis of that robustness \cite{survey-vadillo}. It should be noted that all of the above works are focused on image classification. Adversarial perturbations for NIDS models may not correspond to valid traffic, and even if they do, they may not lead to significant changes in explanation. We evaluate the robustness of the explanation generated for a black-box NIDS model using informal techniques inspired by literature. 

%In the TRUSTEE algorithm, there is an inner loop which is designed to generate different high fidelity DT explanations, one per iteration. It does so by applying a teacher-student dynamic derived from imitation learning that uses the blackbox model, $\pi^*$ as an oracle in conjunction with a carefully curated dataset $D’$ to guide the training of a surrogate ``white-box'' model in the form of a DT that imitates the black-box’s decisions. In contrast, the purpose of the outer loop is (i) to select from among the $N$ high fidelity DTs that have been generated in the process of executing the inner loop the DT with the highest fidelity, (ii) to transform this resulting DT into a high-fidelity and low-complexity DT by means of a post-processing step that consists of applying a purposefully developed pruning method called top-$k$ pruning which carefully scrutinizes only the top $k$ branches of an extracted high-fidelity DT, ranked by the number of input samples a branch classifies, and (iii) to consider all $S$ high-fidelity and low-complexity DTs that have been generated in the process of executing the outer loop and output the one that is the most stable in the sense of having the highest mean agreement among these $S$ DTs.   

\subsubsection{\textbf{Stability}}
The explanations generated for a black-box model using a particular explanation method over multiple runs may not match completely, i.e., some features which are part an explanation generated for one run may not appear in the explanation generated in another run. For a security practitioner, if the feature-based explanation is not \textit{stable} over multiple runs, they may decide against using it. As per \cite{sok-dl-sec}, to determine the stability of an explanation method, we calculate the Intersection Size (IS) for runs $i$ and $j$ as:

\begin{equation}
	IS(i, j) = \frac{|T_i \cap T_j|}{k},
	\label{eq-is}
\end{equation} 
where $T_i$ and $T_j$ are the sets of top $k$ relevant features from the explanation generated in runs $i$ and $j$, respectively. An explanation method is defined as stable if the corresponding intersection size is close to $1$, i.e., $IS(i,j) > 1-\epsilon$ for a small threshold $\epsilon$.   

% Taken from TRUSTEE paper. Rephrased it.
%\subsubsection{\textbf{Inductive Bias}}: A number of ML/DL models face the issue of underspecification. This term refers to the challenge of determining whether a model’s high performance is due to its genuine ability to capture essential patterns in the data or merely the result of certain inductive biases it has learned. These biases often appear as spurious correlations, a lack of ability to adapt to data samples with variable statistics, shortcut learning strategies, etc. The presence of such biases in trained models undermines their reliability and expected performance in real-world applications. Therefore, identifying these inductive biases is crucial for establishing trust in ML/DL models. To detect the presence of an inductive bias in the DL-based NIDSs under consideration in this work, such as vulnerability to variable packet rate samples, we use tampered malicious traffic traces \cite{trustee} which alter the attack segments of the original traffic traces by distributing the malicious packets in such a way that the packet rate remains within a specified limit.

% SHAP implementation issues
% LIME implementation issues
% LEMNA implementation issues

\section{Experimental Results}
\label{results}
In this section, we present explanations of state-of-the-art DL-based NIDSs generated using XAI methods such as TRUSTEE and kernel SHAP. For each pair of NIDS and XAI method, we wrote Python scripts for generating the explanation using the XAI method's API and the modified source code of the NIDS (to make it compatible with the XAI method's API implementation). Using the generated explanations and further analysis, we aim to answer the following research questions:
\begin{enumerate}
%	\item \textit{RQ1: How are the explanations generated for a given NIDS by TRUSTEE and kernel SHAP related to each other?}
%
%We investigate the relation between the explanations generated for a given NIDS by TRUSTEE and kernel SHAP by using the systematic approach for comparing TRUSTEE and SHAP explanations at the DT subset-level as explained in sub-section \ref{trustee-shap-rel}.
	\item \textit{RQ1: How do TRUSTEE and SHAP compare against each other in terms of various evaluation criteria?}

We use the \textit{TRUSTEE-SHAP agreement score} introduced in sub-section \ref{trustee-shap-rel} to investigate the relation between the explanations generated for a given NIDS by TRUSTEE and kernel SHAP. We also use security-focused criteria such as \textit{adversarial robustness} and \textit{stability} to compare TRUSTEE versus SHAP explanations.
	\item \textit{RQ2: What are the most prominent features used by each NIDS model to make decisions on input data samples?}

The explanations generated for each NIDS model by TRUSTEE and kernel SHAP rank the features used for making decisions on input data samples. TRUSTEE ranks the features based on the percentage of total number of test data samples classified using those features while kernel SHAP ranks the features based on the mean absolute SHAP values for those features. 
%	\item \textit{RQ3: Are inductive biases present in the NIDS implementations?}
%
%We investigate the presence of inductive biases (e.g., vulnerability to variable packet rate samples) in the NIDS implementations using tampered malicious traffic traces as explained in sub-section \ref{trustee-shap-rel}.
\end{enumerate}
%Finally, we compare the explanations generated across the four NIDSs by TRUSTEE using the fidelity metric and by kernel SHAP using general XAI evaluation metrics.

\textbf{NIDSs Evaluated}: We have used the state-of-the-art DL-based NIDSs presented in Section \ref{sota-dl-nids} (Kitsune, HorusEeye, ENIDrift and HELAD) for our analysis. When we refer to HorusEye from this point onwards, we are referring to its Magnifier component which performs deep learning-based anomaly detection. Since the feature extraction algorithm code for Deep ResNIDS implementation is not publicly available, we have not included it in our analysis.
%(and there was no response from the authors when we contacted them)
%Since the source code for IoTA NIDS implementation has not been released publicly, we have not included it in our analysis. 

\textbf{Datasets}: We use the Kitsune Mirai dataset \cite{kitsune} and the CICIDS-2017 dataset \cite{cicids-2017} for training/testing the DL-based NIDSs under consideration. The Mirai PCAP trace consists of $\sim$ 760k packets of synthetically generated attack in a network with nine IoT devices, first $\sim$120k packets consist of benign traffic, remaining $\sim$640k packets have anomalous traffic. We use the Kitsune code to extract features from the PCAP trace through the DIS method. The CICIDS-2017 dataset \cite{cicids-2017} consists of five days of packet capture (July 3 to July 7, 2017) from a small-scale enterprise-like network, with the first day consisting of benign background traffic and the other days consisting of attack traffic (brute force SSH, DoS, Heartbleed, web attack, infiltration, etc.). It consists of significant drift over the data collection period in terms of different attack tools used. The CICIDS-2017 dataset provides PCAP traces, labeled network traffic flows and CSV files consisting of pre-selected features extracted from traffic flows. We use the raw PCAP traces directly and extract features from them using the Kitsune DIS method to maintain consistency. Below, we present the explanations generated using Mirai dataset only due to space limitations though we have included a few results obtained using CICIDS-2017 dataset in Section \ref{comp-analysis}. 

\textbf{Test Environment}: To test Kitsune, HorusEye and ENIDrift NIDSs with XAI methods, we used a VMWare ESXi server VM with Intel Xeon Silver 4216 CPU @2.10GHz, 64-bit architecture, 8 cores, 16GB RAM and running Ubuntu 18.04/Ubuntu 20.04 OS. For testing HELAD NIDS, we used an Nvidia RTX 3090 GPU cluster with each GPU @1.4GHz, $\sim$25GB of memory since it requires training deep belief networks and LSTMs. Each GPU had 328 tensor cores and was running Ubuntu 22.04 OS.

\subsection{Comparative Analysis}
\label{comp-analysis}
In this sub-section, we use \textit{XAI evaluation metrics} to quantify the effectiveness of the explanations generated for the DL-based NIDSs under consideration by TRUSTEE and kernel SHAP. TRUSTEE's implementation comes with its own metric called \textit{fidelity} which quantifies how accurate the TRUSTEE DT is with respect to the original black-box ML model. We also use the TRUSTEE-SHAP agreement score proposed in Section \ref{trustee-shap-rel} to compare the level of agreement between TRUSTEE DT and kernel SHAP explanations generated for the DL-based NIDSs considered and security-focused criteria such as adversarial robustness/stability to compare TRUSTEE versus SHAP explanations.
 
\textbf{TRUSTEE Fidelity}: Using the Mirai dataset, if we compare the fidelity values obtained for the NIDSs considered, without any pruning, the highest fidelity values are obtained for Kitsune (0.786) out of the four NIDSs. With top-k pruning ($k=10$), the highest fidelity values are obtained for Kitsune (0.721) out of the four NIDSs. Since ENIDrift, HorusEye and HELAD NIDSs yield negative and zero fidelity values, TRUSTEE DT explanations do not serve much purpose. Since top-k pruning is required to reduce the complexity of DT explanations and make them practical for cyber security personnel, \textit{TRUSTEE is most useful for explaining the decisions of Kitsune NIDS}.

Using the CICIDS-2017 dataset, if we compare the fidelity values obtained for the NIDSs considered without any pruning, the highest fidelity values are obtained for ENIDrift (0.929) out of the four NIDSs. With top-k pruning ($k=10$), the highest fidelity values are obtained for Kitsune (0.919) out of the four NIDSs. Since HorusEye and HELAD NIDSs yield negative and zero fidelity values, TRUSTEE DT explanations do not serve much purpose. With top-k pruning, \textit{TRUSTEE is most useful for explaining the decisions of Kitsune NIDS followed by ENIDrift NIDS.} The comparison of fidelity values obtained during TRUSTEE analysis of the DL-based NIDSs considered in this paper is shown in Table \ref{fidelity-comp-table}.

\begin{tcolorbox}[width=\linewidth, sharp corners=all, colback=white!95!black]
\textbf{Takeaway 1}: Across the datasets, with top-k pruning, Kitsune NIDS yields the highest TRUSTEE DT fidelity values.
\end{tcolorbox}

% try to explain the intuition behind these results
% benchmark against boxplot, single AE
 
%\begin{table}[h]
%	\centering
%%	\vspace*{0.5cm}
%    \begin{tabular}{ | l | l | l | }
%    \hline
%    \textbf{IoT NIDS} & \thead{\textbf{Fidelity}\\ \textbf{(Without pruning)}} & \thead{\textbf{Fidelity (with top-k}\\ \textbf{pruning, k=10)}} \\ \hline
%    Kitsune & 0.786 & 0.721 \\ \hline
%    HorusEye & -1.365 & -1.038e-06 \\ \hline
%    ENIDrift & 0.715 & -1.536e-09 \\ \hline
%    HELAD & 0.0 & 0.0 \\ \hline
%    \end{tabular}
%    \caption{Comparison of TRUSTEE fidelity values with and without pruning for DL-based IoT NIDSs considered}
%    \label{fidelity-comp-table}
%\end{table}

\begin{table}[h]
	\centering
%	\vspace*{0.5cm}
    \begin{tabular}{ | l | l | l | l | l | }
    \hline
    \textbf{NIDS Name} & \multicolumn{2}{|l|}{\thead{\textbf{Fidelity}\\ \textbf{(Without pruning)}}} & \multicolumn{2}{|l|}{\thead{\textbf{Fidelity (with top-k}\\ \textbf{pruning, k=10)}}} \\ \cline{2-5}
	 & \thead{\textbf{Mirai}\\ \textbf{dataset}} & \thead{\textbf{CICIDS-2017}\\ \textbf{dataset}} & \thead{\textbf{Mirai}\\ \textbf{dataset}} & \thead{\textbf{CICIDS-2017}\\ \textbf{dataset}} \\ \hline
    Kitsune & 0.786 & 0.843 & 0.721 & 0.919 \\ \hline
    HorusEye & -1.365 & -0.956 & -1.038e-06 & -0.018 \\ \hline
    ENIDrift & 0.715 & 0.929 & -1.536e-09 & 0.53 \\ \hline
    HELAD & 0.0 & 0.0 & 0.0 & 0.0 \\ \hline
    \end{tabular}
    \caption{Comparison of TRUSTEE fidelity values with and without pruning for DL-based NIDSs considered}
    \label{fidelity-comp-table}
\end{table}

\textbf{SHAP evaluation metrics}: We experimented with multiple implementations of objective XAI evaluation metrics (e.g., completeness, faithfulness, stability, monotonicity) to compare the kernel SHAP explanations across the four NIDSs evaluated in our experiments. Some of those implementations are OpenXAI \footnote{https://github.com/AI4LIFE-GROUP/OpenXAI}, XAI-bench \footnote{https://github.com/abacusai/XAI-bench} and Quantus \footnote{https://github.com/understandable-machine-intelligence-lab/Quantus}. However, after extensive experimentation with those implementations, we found out that none of them support custom ML models such as those used for NIDSs. They support out-of-the-box ML models such as linear regression, MLP and ANN only.

\textbf{Level of TRUSTEE-SHAP Agreement}: We set the value of $m=3$ for the $m$-point average TRUSTEE-SHAP agreement score. Each of the data subsets used for calculation of TRUSTEE-SHAP agreement scores were required to have a minimum number of 300 samples. The results are shown in Table \ref{trustee-shap-comp-table}. HorusEye, with a score of $0.381$, exhibits the highest level of TRUSTEE-SHAP agreement, followed by Kitsune, with a score of $0.167$ while ENIDrift has a score of $0.0$. We also find that the TRUSTEE-SHAP agreement scores obtained for an NIDS for the different data subsets vary significantly. For example, for Kitsune, the scores obtained for three different data subsets were $0.4$, $0.6$ and $0.1$. 

\begin{tcolorbox}[width=\linewidth, sharp corners=all, colback=white!95!black]
\textbf{Takeaway 2}: The level of agreement between TRUSTEE DT explanation and SHAP explanation for a blackbox NIDS model may depend on the specific data subset which is used to conduct the TRUSTEE-SHAP comparison.
\end{tcolorbox}

\begin{table}[h]
	\centering
%	\vspace*{0.5cm}
    \begin{tabular}{ | l | l | l | }
    \hline
    \textbf{NIDS} & \thead{\textbf{$m$-point avg. TRUSTEE-SHAP}\\ \textbf{agreement score (m=3)}} \\ \hline
    Kitsune & 0.167 \\ \hline
    HorusEye & 0.381 \\ \hline  % (2/3 + 2/6 + 1/7)
    ENIDrift & 0.0 \\ \hline
    HELAD & n/a \\ \hline 
    \end{tabular}
    \caption{Comparison of $m$-point average TRUSTEE-SHAP agreement scores for DL-based SOTA NIDSs trained with Kitsune Mirai dataset}
    \label{trustee-shap-comp-table}
\end{table}
 
\textbf{Adversarial Robustness}: SHAP explanations have already been shown to be vulnerable to adversarial input perturbations as per \cite{slack}. Further, global explanation methods such as Partial Dependence (PD) plots have also been subject to adversarial attacks either by poisoning the dataset used to obtain the PD explanation for a biased black-box model to manipulate the explanation \cite{pdp-poison} or using extrapolated permuted PD data \cite{pdp-extrap} to preserve the original black-box model predictions while hiding the model's discriminatory behaviour. There is no existing work on adversarial attacks against TRUSTEE (also a global explanation method), though we surmise possible adversarial attacks which poison the dataset (used to obtain a biased black-box model's explanation) in a way that the curated dataset (sampled from the poisoned dataset) used to guide the surrogate DT model, can be used to manipulate the final DT explanation while maintaining high fidelity. 

\begin{tcolorbox}[width=\linewidth, sharp corners=all, colback=white!95!black]
\textbf{Takeaway 3}: SHAP and TRUSTEE explanations are both vulnerable to adversarial attacks, though further work is needed to quantify their degree of robustness.
\end{tcolorbox}  

\textbf{Stability}: We evaluate the stability scores of TRUSTEE and SHAP explanations for each DL-based NIDS trained over Mirai dataset by calculating the average intersection size (as defined in Equation \ref{eq-is}) of the top $10$ features in the explanations generated by each method over $3$ runs. The scores obtained are shown in Table \ref{stability-score-table}. The average intersection size for ENIDrift is lowest among all the NIDSs for TRUSTEE and SHAP explanations. Across TRUSTEE and SHAP, HorusEye has the highest stability, closely followed by Kitsune.  Overall, SHAP fares slightly better than TRUSTEE in terms of stability. Though kernel SHAP explanations for individual data instances often suffer from instability \cite{shapingshap}, we have used the mean absolute value of the SHAP values for each feature across all the data samples which improves the stability of resulting explanation.

\begin{tcolorbox}[width=\linewidth, sharp corners=all, colback=white!95!black]
\textbf{Takeaway 4}: Across the NIDSs, SHAP explanations yield better stability than the TRUSTEE explanations.
\end{tcolorbox}     

\begin{table}[h]
	\centering
%	\vspace*{0.5cm}
    \begin{tabular}{ | l | l | l | l | l | }
    \hline
    \textbf{Method} & \textbf{Kitsune} & \textbf{HorusEye} & \textbf{ENIDrift} & \textbf{HELAD} \\ \hline
    SHAP & 0.933 & 1.000 & 0.133 & n/a \\ \hline % Kitsune- 9+10+9, Horuseye- 10+10+10, Enidrift- 0+2+2
    TRUSTEE & 0.833 & 1.000 & 0.033 & 0.200 \\ \hline % Kitsune- 8+9+8, Horuseye- 10+10+10, Enidrift- 0+1+0, HELAD- 3+2+1
    \end{tabular}
    \caption{Average intersection size between top $k=10$ features of explanations for 3 runs. Values closer to 1 indicate greater stability.}
    \label{stability-score-table}
\end{table} 

\subsection{Case Study- Kitsune $\times$ Mirai}
\begin{figure}[h]
\centering
\includegraphics[scale=0.35]{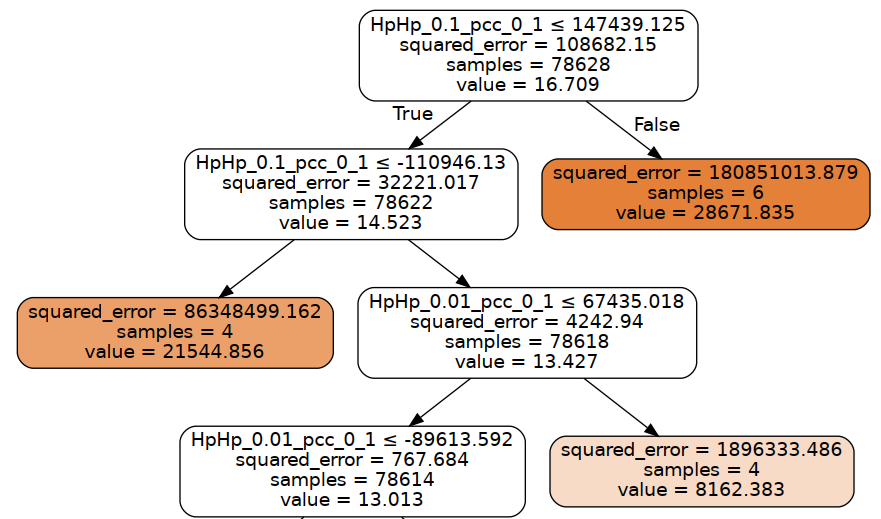}
\caption{Decision tree for Kitsune Mirai model with Top-k pruning (k=3). Only the top 3 layers are shown.}
\label{kitsune-mirai-dt}
\end{figure}

%\subsection{Case Study- Kitsune}
\label{kitsune-case}
\textbf{TRUSTEE Analysis}: The results of TRUSTEE analysis for Kitsune are shown in Table \ref{kitsune-mirai-trustee-table}. The \textit{sample size} refers to the fraction of the training dataset to use to train the student decision tree model. \textit{Fidelity} is defined as the R-squared value between Kitsune’s predictions and those obtained by the DT explanation. Using TRUSTEE with $30\%$ of the original Mirai dataset samples and no pruning results in the DT explanation which achieves 0.786 fidelity compared to Kitsune. Using TRUSTEE with $30\%$ of the dataset samples and top-k pruning method (setting $k = 3$) results in a DT explanation that is shown in Fig. \ref{kitsune-mirai-dt} and achieves 0.697 fidelity compared to Kitsune. We now explain the decision tree briefly: when the feature $HpHp\_0.01\_pcc\_0\_1$ is less than or equal to the value $147439.125$, the control passes to the left branch, whereas when the opposite is true, the control passes to the right branch. \textit{Squared error} refers to the mean-squared error between the target and predicted value for the data points covered in the tree node, \textit{samples} refers to the number of data points covered in the node and \textit{value} refers to the value of the target variable. Dark colored nodes with no feature names represent the leaf nodes.
% explain the decision tree

Using TRUSTEE with $30\%$ of the dataset samples and top-k pruning method (setting $k = 10$) results in a DT explanation which achieves 0.721 fidelity compared to Kitsune. The reduction in fidelity with pruning branches of a DT is expected since the obtained smaller trees run the risk of missing important decision branches as they prevent the consideration of any further decision branches once the stopping criterion is reached.

The fidelity score presents a way for security practitioners to gauge the effectiveness of the explanation generated for an NIDS. A security practitioner prioritising just explainability should select the NIDS with the highest fidelity score. However, a typical practitioner interested in deploying a DL-based NIDS will prioritise not just its explainability but also its its attack detection performance. Traditionally, the detection performance of ML/DL-based NIDSs has been measured in terms of metrics such as accuracy, precision, recall and F1 score. An example of the metric a security practitioner can use while selecting a DL-based NIDS is a weighted combination of the NIDS's average F1 score for different kinds of attacks and TRUSTEE fidelity, expressed as below:

\begin{equation}
	S = \alpha * (F1 \ score) + (1-\alpha) * (Fidelity), 0 \leq \alpha \leq 1
\end{equation}
By varying the value of the weight, $\alpha$, the security practitioner can prioritise between detection performance and fidelity.  
 
Based on the TRUSTEE analysis results obtained earlier, the top three prominent features that Kitsune uses to determine an anomaly are: 
\begin{itemize}
    \item $HpHp\_0.01\_pcc\_0\_1$, $HpHp\_0.1\_pcc\_0\_1$- Correlation coefficient between two packet size streams aggregated by traffic sent between a set of source and destination IP addresses with time windows 0.01 (1 minute) and 0.1 (10 seconds)
    \item $HpHp\_0.1\_weight\_0$- Weights (packet count) aggregated by traffic sent between a set of source and destination IP addresses and associated with the time window 0.1 (1.5 seconds). 
\end{itemize}

\begin{tcolorbox}[width=\linewidth, sharp corners=all, colback=white!95!black]
\textbf{Takeaway 5}: Kitsune's DT relies mainly on the volume of packets and sizes of packets exchanged per time frame to determine if an attack is underway.
\end{tcolorbox}

Information about the top prominent packet-based features used by a DL-based NIDS for anomaly detection makes it easier for a security practitioner interested in deploying the NIDS to trust their internal decision making process. It also enables practitioners to compare NIDSs based on their respective top features. However, practitioners need to select only those NIDSs for comparison whose fidelity is above a certain minimum threshold. If the fidelity falls below that threshold, it means that the DT explanation generated for the corresponding NIDS is not reliable.

\begin{table}[h]
	\centering
%	\vspace*{0.5cm}
    \begin{tabular}{ | l | l | l | l | }
    \hline
    \textbf{Sample size} & \thead{\textbf{Top-k pruning}\\ \textbf{used?}} & \textbf{DT size, depth, leaves} & \textbf{Fidelity} \\ \hline
    30\% & No & 133377, 70, 66689 & 0.786 \\ \hline
    30\% & Yes (k=3) & 19, 9, 10 & 0.697 \\ \hline
    30\% & Yes (k=10) & 37, 13, 19 & 0.721 \\ \hline
    \end{tabular}
    \caption{TRUSTEE analysis results for Kitsune Mirai model}
    \label{kitsune-mirai-trustee-table}
\end{table}

\begin{figure}[h]
\centering
\includegraphics[scale=0.2]{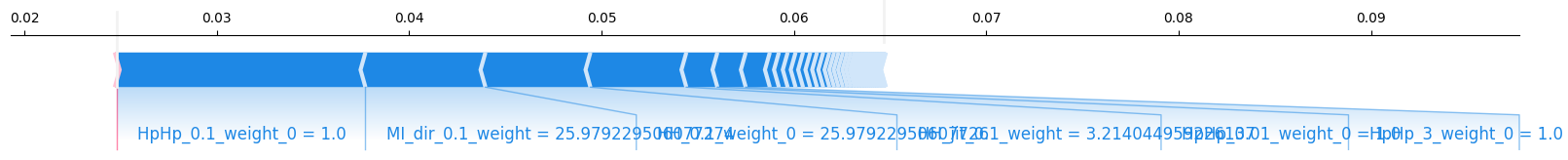}
\caption{SHAP force plot for a single data point from Kitsune Mirai model}
\label{kitsune-mirai-shap}
\end{figure}

\textbf{SHAP Analysis}: Fig. \ref{kitsune-mirai-shap} illustrates a local explanation for a benign packet data point using a SHAP’s force plot for Kitsune, displaying the contribution of each feature to the prediction. The plot shows the base value, and the features containing a positive influence on the prediction are in red, and the features showing a negative influence on the predictions are in blue. The base value in the plots is the average of all prediction values. Each strip in the plot illustrates how the features influence the predicted value, either drawing it closer or pushing it farther away from the base value. Red strip features push the value to higher values, whereas blue strip features push the value to lower values. The contribution of features holding broader strips is more.

The base value is $0.0646$. Features such as $HpHp\_0.1\_weight\_0$ and $MI\_dir\_0.1\_weight$ have a negative impact on the prediction value. $HpHp\_0.1\_weight\_0$ is the most crucial feature, as the contribution has a broader range. The total negative contribution is greater than the positive contribution, and the final predicted value is lesser than the base value. As a result, the class is predicted as benign. The feature $HpHp\_0.1\_weight\_0$ corresponds to weights aggregated by traffic sent between a set of source and destination IP addresses using a specific protocol (ARP/TCP/UDP) with time window 0.1 (10 seconds) while the feature $MI\_dir\_0.1\_weight$ corresponds to weights aggregated by source MAC and IP addresses, with time window 0.1 (10 seconds).

\begin{figure}[h]
\centering
\includegraphics[scale=0.4]{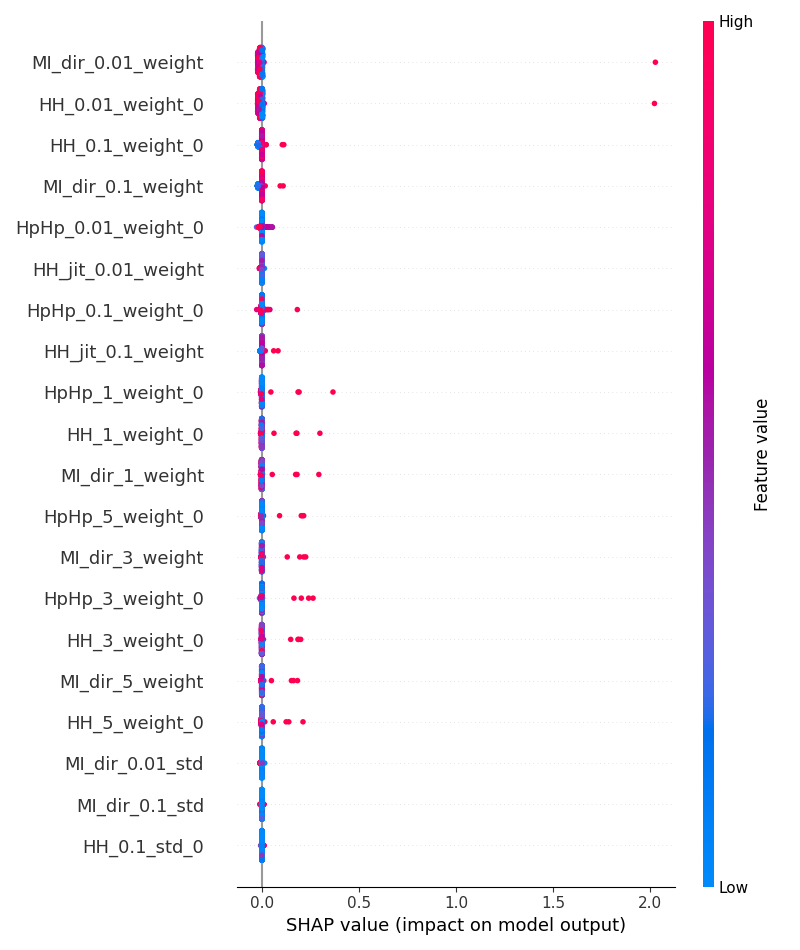}
\caption{SHAP beeswarm plot for Kitsune Mirai model}
\label{kitsune-mirai-shap-beeswarm}
\end{figure}

Using kernel SHAP analysis, Fig. \ref{kitsune-mirai-shap-beeswarm} shows a summary of how the top features in Mirai dataset impact the Kitsune NIDS model’s output through a beeswarm plot for randomly selected 1000 data samples corresponding to the benign root node-leaf node path in the TRUSTEE DT (Fig. \ref{kitsune-mirai-dt}) which takes $625.70$ seconds on an average. Running SHAP analysis for the whole dataset will take substantial amount of time and computing resources. Each instance of the given explanation is represented by a single dot on each feature row. The $x$ position of the dot is determined by the SHAP value of that feature, and dots pile up along each feature row to show density. Color is used to display the original value of a feature. The features are ordered using the mean absolute value of the SHAP values for each feature. Features such as $MI\_dir\_0.1\_weight$, $HH\_0.01\_weight\_0$ and $HH\_0.1\_weight\_0$ have the most impact on the prediction value, in that order. $HH\_0.01\_weight\_0$ and $HH\_0.1\_weight\_0$ correspond to weights aggregated by traffic sent between a set of source and destination IP addresses with time windows 0.01 (1 minute) and 0.1 (10 seconds) respectively.

\subsection{Case Study- HorusEye $\times$ Mirai}
%\subsection{Case Study- HorusEye}
\label{horuseye-case}
%We used the dataset provided by HorusEye \cite{horuseye} authors who randomly down-sampled the Mirai attack traffic in Kitsune dataset to consist of 40,000 packets. The frequency of Mirai attacks is too high in the original dataset to reflect real botnet infection.

\textbf{TRUSTEE Analysis}: The results of TRUSTEE analysis for HorusEye are shown in Table \ref{horuseye-mirai-trustee-table}. Using TRUSTEE with $30\%$ of the Mirai dataset samples with no pruning results in a DT explanation which achieves negative fidelity ($\sim -1.365$) compared to original HorusEye model. Using TRUSTEE with $30\%$ of the dataset samples and top-k Pruning method (setting $k = 10$) results in a DT explanation which achieves negative fidelity ($\sim -1.038e-06$) compared to original HorusEye model. Negative fidelity values mean that the DT explanation fits HorusEye predictions worse than the mean of the predictions. 

\begin{figure}[h]
\centering
\includegraphics[scale=0.35]{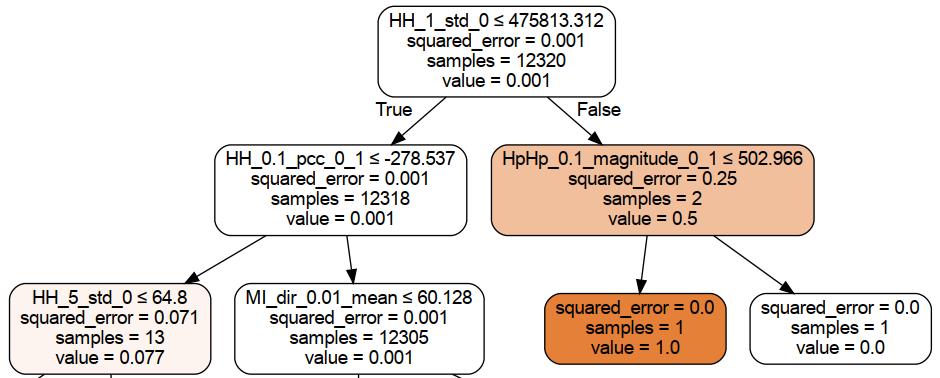}
\caption{Decision tree for HorusEye Mirai model with Top-k pruning (k=10). Only the top 3 layers are shown.}
\label{horuseye-mirai-dt}
\end{figure}

Nevertheless, the decision tree for HorusEye Mirai model with top-k pruning (k=10) is shown in Fig. \ref{horuseye-mirai-dt}. The top three prominent features that HorusEye's DT uses to determine an anomaly are:
\begin{enumerate}
    \item $HH\_3\_pcc\_0\_1$- Correlation coefficient between two packet size streams aggregated by traffic sent between a set of source and destination IP addresses with time window 3 (500 milli-seconds).
    \item $HH\_1\_std\_0$- Standard deviation of packet sizes aggregated by the traffic sent between a set of source and destination IP addresses with time window 1 (1.5 seconds).
    \item $MI\_dir\_0.01\_mean$- Mean of packet sizes aggregated by source MAC and IP addresses, with time window 0.01 (1 minute).
\end{enumerate}

\begin{tcolorbox}[width=\linewidth, sharp corners=all, colback=white!95!black]
\textbf{Takeaway 6}: HorusEye's DT relies mainly on the sizes of packets exchanged between a pair of IP addresses per time frame to determine if an attack is underway.
\end{tcolorbox}

\begin{table}[h]
	\centering
%	\vspace*{0.5cm}
    \begin{tabular}{ | l | l | l | l | }
    \hline
    \textbf{Sample size} & \thead{\textbf{Top-k pruning}\\ \textbf{used?}} & \textbf{DT size, depth, leaves} & \textbf{Fidelity} \\ \hline
    30\% & No & 45, 11, 23 & -1.365 \\ \hline
    30\% & Yes (k=10) & 1, 0, 1 & -1.038e-06 \\ \hline
    \end{tabular}
    \caption{TRUSTEE analysis results for HorusEye Mirai model}
    \label{horuseye-mirai-trustee-table}
\end{table}

%\textbf{SHAP Analysis}: In Fig. \ref{horuseye-shap}, we show the local explanation for a benign packet data point using the SHAP’s force plot for HorusEye. The base value is $0.0$. The feature $HpHp\_5\_mean\_0$ has a negative impact on the prediction value while features such as $HH\_5\_magnitude\_0\_1$ and $HpHp\_3\_covariance\_0\_1$ have a positive impact. $HH\_jit\_3\_mean$ is the most crucial feature, as the contribution has a broader range. The total negative contribution is equal to the positive contribution, and the final predicted value is equal to the base value. As a result, the class is predicted as benign. 

%Here, the feature $HpHp\_5\_mean\_0$ corresponds to the mean of packet sizes aggregated by traffic sent between a set of source and destination IP addresses with time window 5 (100 ms), the feature $HH\_5\_magnitude\_0\_1$ corresponds to the \textit{L2}-norm of the means of two packet size streams aggregated by traffic sent between a set of source and destination IP addresses with time window 5 (100 ms) and the feature $HpHp\_3\_covariance\_0\_1$ corresponds to the covariance between two packet size streams aggregated by traffic sent between a set of source and destination IP addresses with time window 3 (500 ms). 

\begin{figure}[h]
\centering
\includegraphics[scale=0.4]{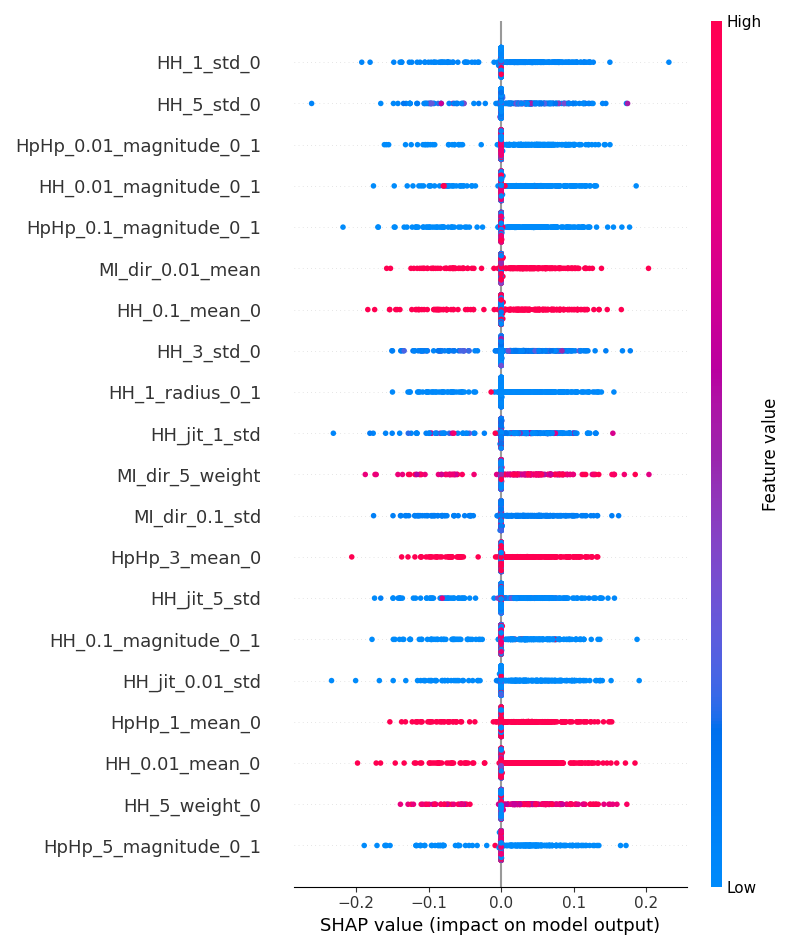}
\caption{SHAP beeswarm plot for HorusEye Mirai model}
\label{horuseye-mirai-shap-beeswarm}
\end{figure}

\textbf{SHAP Analysis}: Using kernel SHAP analysis, Fig. \ref{horuseye-mirai-shap-beeswarm} shows a summary of how the top features in Mirai dataset impact the HorusEye NIDS model’s output using a beeswarm plot for randomly selected 1000 data samples corresponding to the benign root node-leaf node path in the TRUSTEE DT (Fig. \ref{horuseye-mirai-dt}). Features such as $HH\_1\_std\_0$, $HH\_5\_std\_0$ and $HpHp\_0.01\_magnitude\_0\_1$ have the most impact on the prediction value, in that order. $HH\_1\_std\_0$ and $HH\_5\_std\_0$ correspond to the standard deviation of packet sizes aggregated by the traffic sent between a set of source and destination IP addresses with time windows 1 (1.5 seconds) and 5 (100 ms) respectively, while $HpHp\_0.01\_magnitude\_0\_1$ corresponds to the absolute magnitude of two packet size streams aggregated by the traffic sent between a set of source and destination IP addresses with time window 0.01 (1 minute).

%Fig. \ref{horuseye-shap-beeswarm-trustee} shows the Shapley values for features shown in the major benign path for HorusEye's DT trained with the downsampled Mirai dataset using a SHAP’s beeswarm plot. Features such as $HpHp\_0.01\_pcc\_0\_1$, $HH\_1\_pcc\_0\_1$ and $HH\_3\_radius\_0\_1$ have the most impact on the prediction value, in that order. The feature $HH\_0.1\_weight\_0$ corresponds to weights aggregated by traffic sent between a set of source and destination IP addresses with time window 0.1 (10 seconds). It can be inferred from this figure that most Shapley values are negative, meaning that Shapley local interpretation agrees with TRUSTEE that, for the selected data samples, these six features’ values are pushing them toward being benign.

\subsection{Case Study- ENIDrift $\times$ Mirai}
%\subsection{Case Study- ENIDrift}
\label{enidrift-case}
%Change the results to AE ensemble instead of PCA ensemble after journal review.

\begin{figure}[h]
\centering
\includegraphics[scale=0.5]{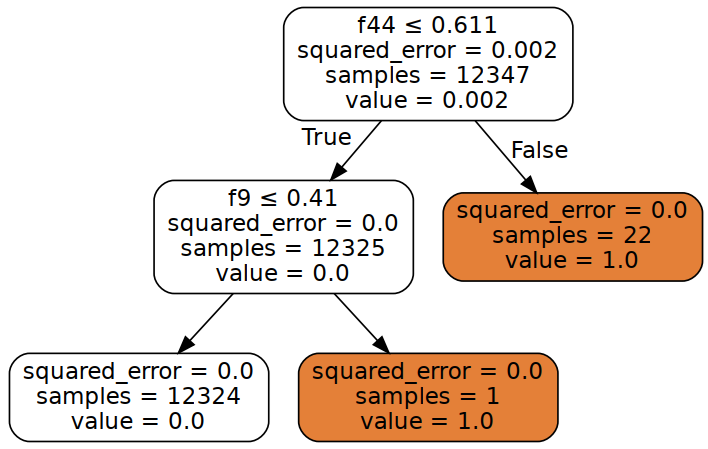}
\caption{Decision tree for ENIDrift Mirai model with Top-k pruning (k=10).}
\label{enidrift-mirai-dt}
\end{figure}

\textbf{TRUSTEE Analysis}: The results of TRUSTEE analysis for ENIDrift are shown in Table \ref{enidrift-mirai-trustee-table}. Using TRUSTEE with $30\%$ of the Mirai dataset samples and no pruning results in the DT explanation which achieves 0.715 fidelity compared to ENIDrift. Using TRUSTEE with $30\%$ of the dataset samples and top-k Pruning method (setting $k = 10$) results in a DT explanation (shown in Fig. \ref{enidrift-mirai-dt}) which achieves negative fidelity ($\sim -1.536e-09$) compared to ENIDrift. This is expected since pruning the branches of a DT reduces its fidelity. The most prominent features ENIDrift uses to determine anomaly are $f44$ and $f9$. \textit{ENIDrift does not extract explicit features from incoming packets. Instead, it uses \textit{ip2V} incremental embedding technique based to embed packets to vectors.} The embedding vector length in ENIDrift implementation is 200.

\begin{table}[h]
	\centering
%	\vspace*{0.5cm}
    \begin{tabular}{ | l | l | l | l | }
    \hline
    \textbf{Sample size} & \thead{\textbf{Top-k pruning}\\ \textbf{used?}} & \textbf{DT size, depth, leaves} & \textbf{Fidelity} \\ \hline
    30\% & No & 5, 2, 3 & 0.715 \\ \hline
    30\% & Yes (k=10) & 1, 0, 1 & -1.536e-09 \\ \hline
    \end{tabular}
    \caption{TRUSTEE analysis results for ENIDrift Mirai model }
    \label{enidrift-mirai-trustee-table}
\end{table}

%\textbf{SHAP Analysis}: In Fig. \ref{enidrift-shap-beeswarm}, we show the local explanation for a malicious packet data point using the SHAP’s force plot for ENIDrift. The base value is $0.925$. The total positive contribution is greater than the negative contribution, and the final predicted value is greater than the base value. As a result, the class is predicted as malicious. A number of features have a positive and negative impacts on the prediction value. However, the kernel SHAP force plot does not show the feature names clearly, perhaps due to the large number of features having similar positive contributions and similar negative contributions to the prediction value. Again, we would like to point out that ENIDrift does not extract explicit features from incoming packets, instead it embeds them to vectors. So, the kernel SHAP force plot is in fact, showing the contributions of embedding vector components.

\begin{figure}[h]
\centering
\includegraphics[scale=0.4]{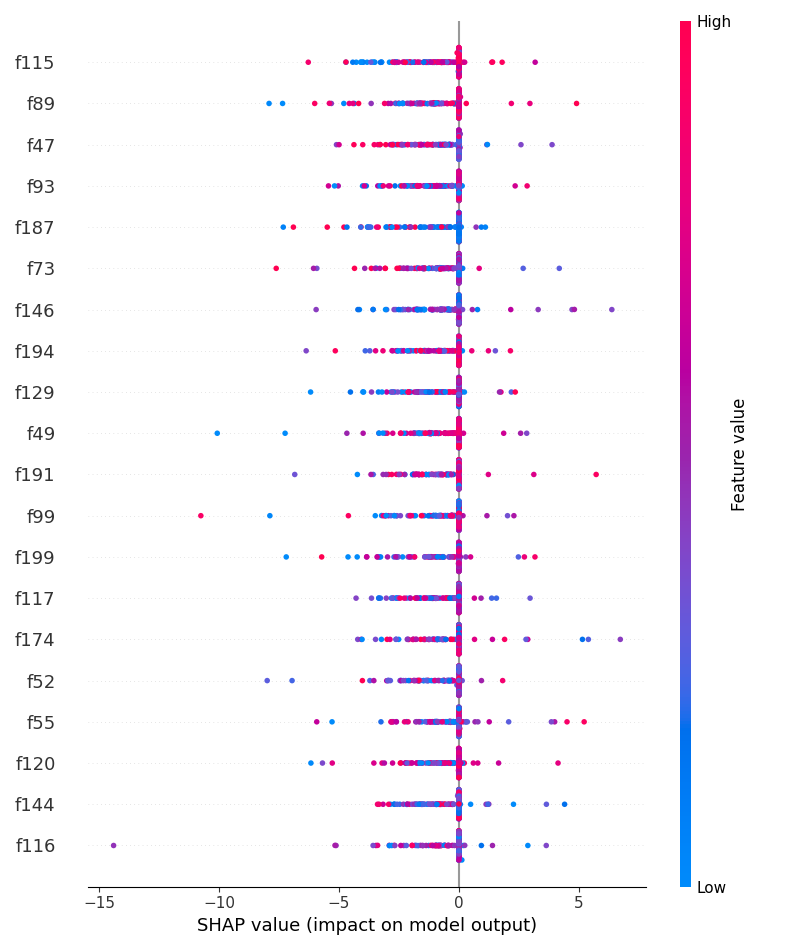}
\caption{SHAP beeswarm plot for ENIDrift Mirai model}
\label{enidrift-mirai-shap-beeswarm}
\end{figure}

\textbf{SHAP Analysis}: Fig. \ref{enidrift-mirai-shap-beeswarm} shows a summary of how the top features in Mirai dataset impact the ENIDrift NIDS model’s output using a beeswarm plot for randomly selected 300 data samples. Features (embedding vector components) such as $f115$, $f89$, $f47$ have the most impact on the prediction value, in that order.

\subsection{Case Study- HELAD $\times$ Mirai}
%\subsection{Case Study- HELAD}
\label{helad-case}

\begin{figure}[h]
\centering
\includegraphics[scale=0.35]{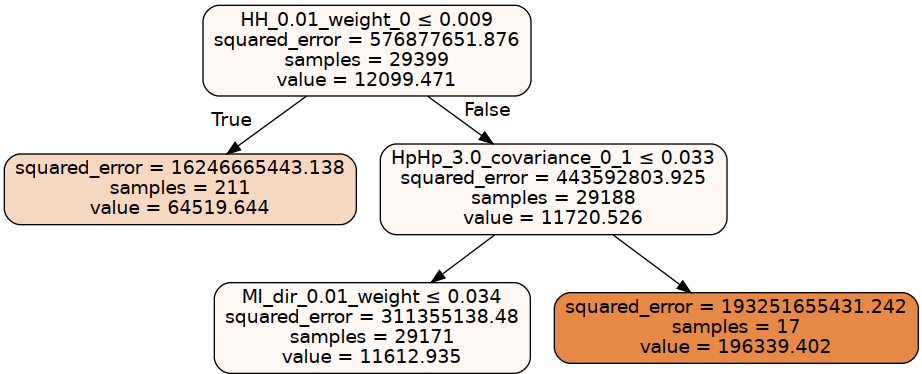}
\caption{Decision tree for HELAD Mirai model with Top-k pruning (k=10). Only the top 3 layers are shown.}
\label{helad-mirai-dt}
\end{figure}

\textbf{TRUSTEE Analysis}: The results of TRUSTEE analysis for HELAD are shown in Table \ref{helad-mirai-trustee-table}. Using TRUSTEE with $30\%$ of the Mirai dataset samples and no pruning results in the DT explanation which achieves 0.0 fidelity compared to HELAD. Using TRUSTEE with $30\%$ of the dataset samples and top-k Pruning method (setting $k = 10$) results in a DT explanation (shown in Fig. \ref{helad-mirai-dt}) which achieves 0.0 fidelity compared to HELAD. A fidelity value of 0.0 means that the DT explanation fits HELAD predictions only as good as the mean of the predictions.

The top three prominent features HELAD's DT uses to determine an anomaly are:
\begin{enumerate}
	\item $MI\_dir\_3.0\_weight$- Weights aggregated by source MAC and IP addresses, with time window 3.0 (500 milli-seconds).
	\item $HH\_1.0\_magnitude\_0\_1$- Absolute magnitude of two packet size streams aggregated by the traffic sent between a set of source and destination IP addresses with time window 1.0 (1.5 seconds).
    \item $HH\_0.01\_weight\_0$- Weights aggregated by traffic sent between a set of source and destination IP addresses with time window 0.01 (1 minute).
\end{enumerate}

\begin{tcolorbox}[width=\linewidth, sharp corners=all, colback=white!95!black]
\textbf{Takeaway 7}: HELAD's DT relies mainly on the volume of packets and the sizes of packets exchanged between a pair of IP addresses per time frame to determine if an attack is underway.
\end{tcolorbox}

\begin{table}[h]
	\centering
%	\vspace*{0.5cm}
    \begin{tabular}{ | l | l | l | l | }
    \hline
    \textbf{Sample size} & \thead{\textbf{Top-k pruning}\\ \textbf{used?}} & \textbf{DT size, depth, leaves} & \textbf{Fidelity} \\ \hline
    30\% & No & 26289, 100, 13145 & 0.0 \\ \hline
    30\% & Yes (k=10) & 435, 94, 218 & 0.0 \\ \hline
    \end{tabular}
    \caption{TRUSTEE analysis results for HELAD Mirai model}
    \label{helad-mirai-trustee-table}
\end{table}

\textbf{SHAP Analysis}: Running a kernel SHAP analysis of HELAD NIDS with the Mirai dataset resulted in issues. First, we ran the analysis with 200,000 packets but it threw a \textit{numpy} memory error since it was unable to allocate a huge amount of memory (335 GiB) for storing a ($2248 \times 20000000$) array with 64-bit float values. Next, we ran the analysis with 20,000 packets but the LSTM was taking too long to train. Hence, we decided to not proceed with the kernel SHAP analysis.

\section{Discussion and Future Work}
\label{discuss}
From the experimental results obtained in the previous section, we have seen that \textit{some DL-based NIDS models (Kitsune, ENIDrift) can be better interpreted in terms of decision trees than other models (HorusEye, HELAD)}. This might be attributed to the complexity of the NIDS model architecture and the properties of the model components. For example, HorusEye NIDS fidelity values are in the negative. Though HorusEye architecture includes only a single autoencoder in the second stage (which processes control plane traffic) compared to an ensemble of AEs in Kitsune or ENIDrift, that single AE is modified from a conventional AE to keep it lightweight, reduce the computational complexity, reduce the false-positive rate and improve throughput. It uses an asymmetric AE with separable convolution, dilation convolution and model quantization. As it is, AEs are limited in their interpretability as a result of their non-linear nature. Due to the complex architecture of the modified AE deployed in HorusEye, TRUSTEE might not be able to build a DT which can imitate the decisions of the black-box modified AE within an acceptable error rate. Therefore, to build DL-based NIDSs with better ``explainability'', we recommend using less complex DL architectures and conventional components. However, there will be a trade-off between ``explainability'' and model performance (e.g., false-positive rate) and NIDS designers need to achieve a balance between the two. One of the ways to achieve this balance is to use the weighted metric introduced in Section \ref{kitsune-case}.

We have also seen from the results that \textit{TRUSTEE and SHAP explanations are in conflict for 2 out of 3 DL-based NIDSs considered in our work}. Their explanations are in agreement for HorusEye NIDS but in conflict for Kitsune and ENIDrift NIDSs. This is probably due to the fact that SHAP value for each feature is calculated for a subset of data samples and reflects the importance of that feature in the decision made by the black-box model for a majority of the data samples in that subset. On the other hand, TRUSTEE acts on the complete input dataset and identifies the top features contributing the most to forming the decision boundary separating all benign and malicious data samples which are part of the input dataset. Another reason could be the different optimization criteria used by TRUSTEE and SHAP while building explanations. TRUSTEE applies a teacher-student dynamic derived from imitation learning that uses the black-box ML/DL model as an oracle in conjunction with a carefully curated dataset to guide the training of a surrogate ``white-box'' model in the form of a DT that imitates the black-box’s decisions. On the other hand, SHAP uses the Shapley kernel method to approximate the black-box model-interpretable model fidelity function which recovers Shapley values. The specific data subset used to conduct the TRUSTEE-SHAP comparison could also play a role in the conflict. To reduce the conflict between TRUSTEE and SHAP explanations at the subset-level, we need to further investigate its underlying cause in the future.
 
Zhao et al. \cite{zhao} have shown that explanations can leak information about the black-box model. This information can be used for model inversion attacks which reconstruct input data using corresponding predictions and model explanation. The robustness of explanation techniques to model inversion attacks is an important criterion, especially when it comes to NIDSs. This is because if an NIDS model can be reconstructed, it can be used by attackers to craft attack traffic to evade detection by that NIDS. However, there exist techniques \cite{jeong}, limited as they may be, to generate inversion-resistant model explanations. Again, it should be noted that all of the above works are focused on image classification. Further work is needed to assess the robustness of explanation methods such as SHAP and TRUSTEE to being exploited for black-box model inversion.

\section{Related Work}
\label{literature}
%\subsection{Need For Explanation of DL-based IoT NIDSs}
Several works have recommended employing explanation techniques for ML systems deployed for cybersecurity. In \cite{dos-n-donts-ml}, the authors have recommended employing explanation techniques to delve deeper into the features of learning-based security systems. Not withstanding their limitations, these techniques can unveil spurious correlations and empower experts to evaluate their effect on the security system's features. The authors in \cite{ml-insec} have also emphasized the importance of explaining the outcomes produced by ML-based detectors used in security contexts. Beyond mere predictions, a cybersecurity threat detection model can offer valuable insights. By understanding the model’s explanations, security practitioners can enhance protection around monitored assets in subsequent executions. Explainability varies in relevance across different domains, but the authors contend that it is crucial for several cybersecurity tasks, enabling the effective application of countermeasures against security threats. The participants (security practitioners) of a survey conducted by the authors of \cite{pragmatic-ml} have expressed the view that security solution providers ought to prioritize methods that offer clear explanations to their clients. 

%\subsection{Evaluation of Explanation Methods on DL-based Security Models}
As a result, we can see the beginning of efforts in the research community to apply XAI techniques to ML-based security systems. A few works \cite{kalakoti, patil, arreche, barnard} have employed XAI methods such as LIME \cite{lime} and SHAP \cite{shap} to assess the quality of local/global explanations generated for conventional machine learning classifiers (AdaBoost, k-nearest neighbour, Multi-layer Perceptron, Random Forest, XGBoost, Light Gradient-Boosting Machine, Gradient Boosting Classifier) trained on network intrusion datasets (NSL-KDD dataset \cite{nsl-kdd}, CICIDS-2017 \cite{cicids-2017}) in terms of criteria such as faithfulness, stability, complexity, and sensitivity. In \cite{warnecke}, the authors have delved into six explanation methods, evaluating their effectiveness across four security systems described in existing literature. These systems leverage DL techniques to detect Android malware, malicious PDF files and security vulnerabilities. The assessment criteria encompass both general properties of deep learning and domain-specific aspects relevant to security. 

The authors in \cite{maonan} have introduced a framework rooted in SHAP to provide explanations for IDSs. This framework blends both local and global explanations, enhancing the overall interpretability of IDSs. Local explanations shed light on why the ML model employed by a specific IDS reaches particular decisions for individual inputs. Meanwhile, global explanations highlight crucial features extracted from the ML model and elucidate the relationships between feature values and specific attack types. \cite{shtayat} has presented a deep learning-based IDS for IIoT networks consisting of an ensemble of three CNN models and an extreme-learning machine model. The authors have used SHAP and LIME techniques to explain the ensemble detector’s decision-making process.
%In \cite{qingtian}, the authors have reviewed the application of XAI methods such as TRUSTEE \cite{trustee} and SHAP to neural network-based Domain Name System (DNS) cache poisoning detection and focused on the connections between network-level explanations and per-data-sample interpretations.
% Include USENIX SEC'23 paper on xNIDS

%\subsection{Explainable DL-based Attack Detection}
A few works have also proposed ML/DL-based attack detection systems with built-in explainability. In \cite{doh-explain}, the authors have implemented an XAI solution to provide accurate detection and classification of the DNS-over-HTTPS attacks. It is based on a balanced stacked random forest classifier. The authors have highlighted the underlying feature contributions to provide transparent and explainable results from the model. In \cite{ensemble-explain}, the authors have proposed an NIDS for IT networks combining ensemble learning and stacking with a meta-learner (CNN) that works on graphical representation of traffic flows and provides the required explainability level for the decisions made. They also provide visual representations of network anomalies that allows security analysts to interpret and gain insights into the detected network anomalies.

Though not targeted at generating explanations, recent works on adversarial learning in NIDSs aim to generate specific examples/feature vectors which can evade detection. Based on a complete/partial knowledge of the ML model underlying a given NIDS and the features used to the train the model, researchers have used existing techniques such as genetic algorithm/particle swarm optimization/generative adversarial networks to generate adversarial examples \cite{bastian-1}. An update to that work focuses on restricting the adversarial feature space so that the generated features correspond to functional network packets \cite{bastian-2}. It does that by presenting a constrained optimization formulation for perturbing raw packet payloads while avoiding modification to the original packet function. A meta-heuristic inspired by genetic algorithm is proposed to solve the optimization problem. However, these works do not offer an explanation as to why the adversarial examples generated cause mis-classification by the underlying ML model.

Despite the presence of existing works on assessing the explainability of ML/DL and ensemble learning-based NIDSs, our work fills an important gap due to the following reasons:
\begin{itemize}
    \item Existing works that evaluate XAI methods on ML/DL-based NIDSs (\cite{kalakoti, patil, arreche, barnard, shtayat, ensemble-explain}): (1) do not compare the explainability of those NIDSs with that of state-of-the-art DL-based NIDSs and (2) do not include the latest XAI tools for ML/DL-based network security such as TRUSTEE \cite{trustee}.
    \item Existing explainable-by-design AI-based attack/intrusion detection systems such as \cite{doh-explain} are targeted at specific attacks (DNS-over-HTTPS attacks). 
\end{itemize}

\section{Conclusion}
\label{conclusion}
We have analyzed four state-of-the-art DL-based NIDS models (Kitsune, HorusEye, ENIDRift, HELAD) using XAI techniques such as TRUSTEE and SHAP. We have compared the explanations generated across XAI methods using our proposed TRUSTEE-SHAP agreement score as well as other security-focused criteria. Using both global explanations and local explanations for the blackbox models' decisions, we have presented the most prominent features used by each NIDS model considered. The results show that: (1) some DL-based NIDS models can be better interpreted in terms of decision trees than other models, (2) TRUSTEE and SHAP explanations are in conflict for most of the NIDS models considered in this work and (3) both TRUSTEE and SHAP explanations are vulnerable to adversarial attacks though TRUSTEE provides better stability than SHAP.

In the future, we plan to design an NIDS using DL architectures of lower complexity and compare its ``explainability'' with other state-of-the-art NIDSs using appropriate metrics. We also plan to investigate the reason(s) behind the lack of agreement between TRUSTEE and SHAP explanations for an NIDS at the data subset-level and use that information to improve the level of agreement.

%\section*{Acknowledgment}
%The authors would like to appreciate the National Cybersecurity R\&D Lab, Singapore for allowing us to use their testbed to collect important data which has been used in our work. This research is supported by the National Research Foundation, Prime Minister’s Office, Singapore under its Corporate Laboratory@University Scheme, National University of Singapore, and Singapore Telecommunications Ltd.

%\clearpage
\bibliographystyle{elsarticle-num}
%\begingroup
%\raggedright
\bibliography{XAIbib}
%\endgroup

\appendix

\section{More Experimental Results}
\label{more-results}
In this section, we present explanations of the four state-of-the-art DL-based NIDSs (Kitsune, HorusEeye, ENIDrift, HELAD) generated using XAI methods such as TRUSTEE and kernel SHAP where the NIDSs were trained/tested with the CICIDS-2017 dataset.

\subsection{Case Study- Kitsune $\times$ CICIDS-2017}
\begin{figure}[h]
\centering
\includegraphics[scale=0.35]{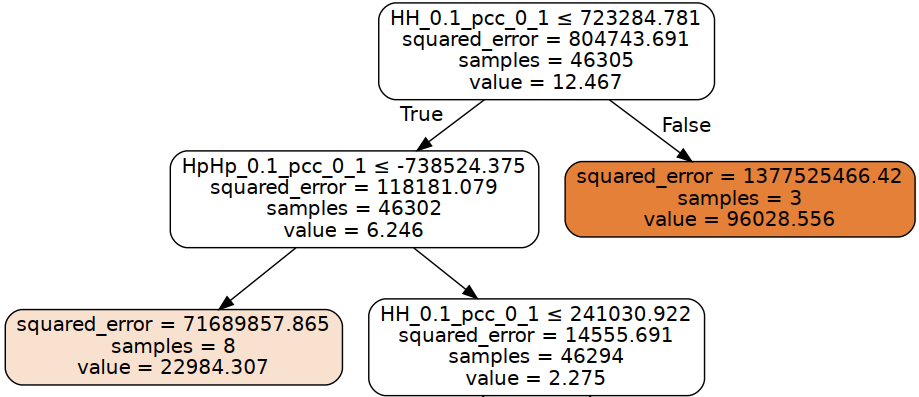}
\caption{Decision tree for Kitsune CICIDS-2017 model with Top-k pruning (k=3). Only the top 3 layers are shown.}
\label{kitsune-cicids-dt}
\end{figure}

%\subsection{Case Study- Kitsune}
\label{kitsune-cicids17-case}
\textbf{TRUSTEE Analysis}: The results of TRUSTEE analysis for Kitsune are shown in Table \ref{kitsune-cicids-trustee-table}. Using TRUSTEE with $30\%$ of the original CICIDS-2017 dataset samples and no pruning results in the DT explanation which achieves 0.843 fidelity compared to Kitsune. Using TRUSTEE with $30\%$ of the dataset samples and top-k Pruning method (setting $k = 3$) results in a DT explanation that is shown in Fig. \ref{kitsune-cicids-dt} and achieves 0.919 fidelity compared to Kitsune.
 
The top three prominent features that Kitsune uses to determine an anomaly are: $HpHp\_0.1\_pcc\_0\_1$, $HH\_0.1\_pcc\_0\_1$ and $HH\_jit\_0.01\_std$. \textit{Thus, Kitsune relies mainly on the sizes of packets and the jitter experienced by packets exchanged per time frame to determine if an attack is underway.}

\begin{table}[h]
	\centering
%	\vspace*{0.5cm}
    \begin{tabular}{ | l | l | l | l | }
    \hline
    \textbf{Sample size} & \thead{\textbf{Top-k pruning}\\ \textbf{used?}} & \textbf{DT size, depth, leaves} & \textbf{Fidelity} \\ \hline
    30\% & No & 78819, 80, 39410 & 0.843 \\ \hline
    30\% & Yes (k=3) & 29, 14, 15 & 0.919 \\ \hline
    \end{tabular}
    \caption{TRUSTEE analysis results for Kitsune CICIDS-2017 model}
    \label{kitsune-cicids-trustee-table}
\end{table}

\begin{figure}[h]
\centering
\includegraphics[scale=0.4]{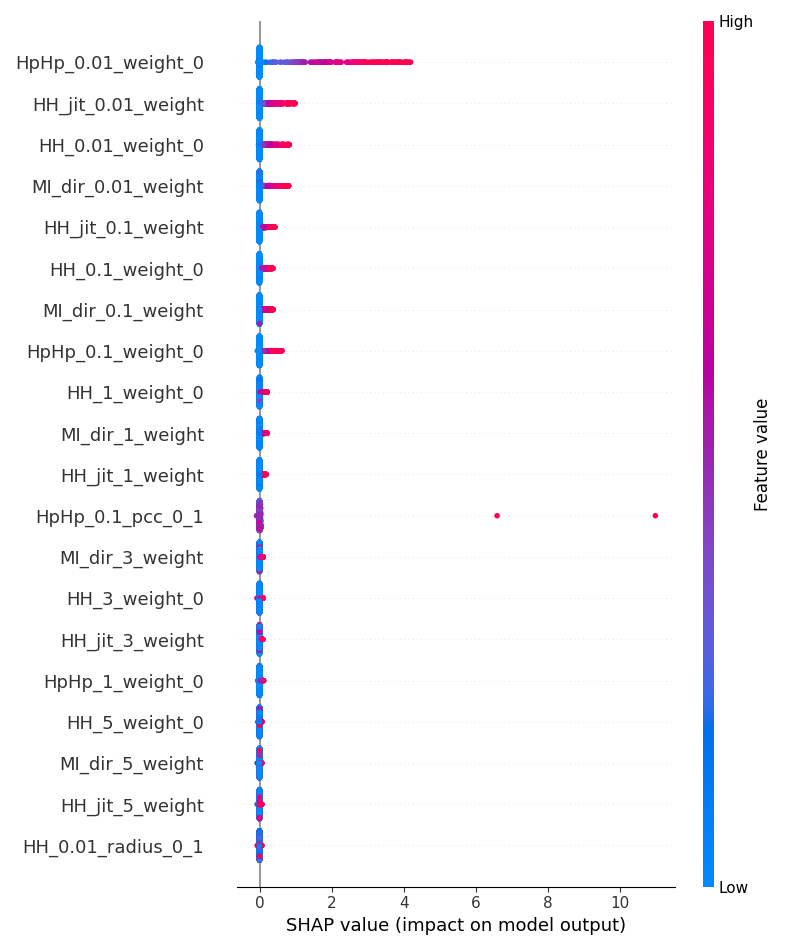}
\caption{SHAP beeswarm plot for Kitsune CICIDS-2017 model}
\label{kitsune-cicids-shap-beeswarm}
\end{figure}

\textbf{SHAP Analysis}: Using kernel SHAP analysis, Fig. \ref{kitsune-cicids-shap-beeswarm} shows a summary of how the top features in CICIDS-2017 dataset impact the Kitsune NIDS model’s output through a beeswarm plot for randomly selected 1000 data samples corresponding to the benign root node-leaf node path in the TRUSTEE DT (Fig. \ref{kitsune-cicids-dt}). Features such as $HpHp\_0.01\_weight\_0$, $HH\_jit\_0.01\_weight$ and $HH\_0.01\_weight\_0$ have the most impact on the prediction value, in that order.

%\textbf{TRUSTEE vs SHAP Analysis}: Using the systematic approach for comparing TRUSTEE and SHAP explanations as explained in sub-section \ref{trustee-shap-rel}, since the set of features used in the benign root node-leaf node path in the TRUSTEE DT is not a subset of the set of top contributing features computed by kernel SHAP summary analysis ($HpHp\_0.1\_pcc\_0\_1$ is the only common feature between the two sets), \textit{TRUSTEE AND SHAP explanations do not agree with each other at the subset level for Kitsune NIDS.}

\subsection{Case Study- HorusEye $\times$ CICIDS-2017}
%\subsection{Case Study- HorusEye}
\label{horuseye-case}
%We used the dataset provided by HorusEye \cite{horuseye} authors who randomly down-sampled the Mirai attack traffic in Kitsune dataset to consist of 40,000 packets. The frequency of Mirai attacks is too high in the original dataset to reflect real botnet infection.

\textbf{TRUSTEE Analysis}: The results of TRUSTEE analysis for HorusEye are shown in Table \ref{horuseye-cicids-trustee-table}. Using TRUSTEE with $30\%$ of the CICIDS-2017 dataset samples with no pruning results in a DT explanation which achieves negative fidelity ($\sim -0.956$) compared to original HorusEye model. Using TRUSTEE with $30\%$ of the dataset samples and top-k Pruning method (setting $k = 10$) results in a DT explanation which achieves negative fidelity ($\sim -0.018$) compared to original HorusEye model but higher fidelity compared to no pruning. Negative fidelity values mean that the DT explanation fits HorusEye predictions worse than the mean of the predictions. 

\begin{figure}[h]
\centering
\includegraphics[scale=0.35]{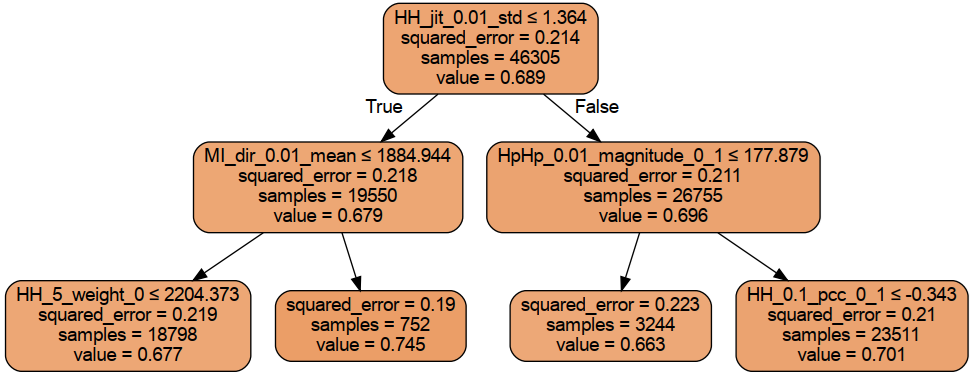}
\caption{Decision tree for HorusEye CICIDS-2017 model with Top-k pruning (k=10). Only the top 3 layers are shown.}
\label{horuseye-cicids-dt}
\end{figure}

Nevertheless, the decision tree for HorusEye Mirai model with top-k pruning (k=10) is shown in Fig. \ref{horuseye-cicids-dt}. The top three prominent features that HorusEye's DT uses to determine an anomaly are: $HH\_5\_pcc\_0\_1$, $HpHp\_1\_covariance\_0\_1$ and $HH\_jit\_0.01\_std$. \textit{Thus, HorusEye's DT relies mainly on the sizes of packets and the jitter experienced by packets exchanged between a pair of IP addresses per time frame to determine if an attack is underway.}

\begin{table}[h]
	\centering
%	\vspace*{0.5cm}
    \begin{tabular}{ | l | l | l | l | }
    \hline
    \textbf{Sample size} & \thead{\textbf{Top-k pruning}\\ \textbf{used?}} & \textbf{DT size, depth, leaves} & \textbf{Fidelity} \\ \hline
    30\% & No & 20265, 94, 10133 & -0.956 \\ \hline
    30\% & Yes (k=10) & 241, 52, 121 & -0.018 \\ \hline
    \end{tabular}
    \caption{TRUSTEE analysis results for HorusEye CICIDS-2017 model}
    \label{horuseye-cicids-trustee-table}
\end{table}

%\textbf{SHAP Analysis}: In Fig. \ref{horuseye-shap}, we show the local explanation for a benign packet data point using the SHAP’s force plot for HorusEye. The base value is $0.0$. The feature $HpHp\_5\_mean\_0$ has a negative impact on the prediction value while features such as $HH\_5\_magnitude\_0\_1$ and $HpHp\_3\_covariance\_0\_1$ have a positive impact. $HH\_jit\_3\_mean$ is the most crucial feature, as the contribution has a broader range. The total negative contribution is equal to the positive contribution, and the final predicted value is equal to the base value. As a result, the class is predicted as benign. 

%Here, the feature $HpHp\_5\_mean\_0$ corresponds to the mean of packet sizes aggregated by traffic sent between a set of source and destination IP addresses with time window 5 (100 ms), the feature $HH\_5\_magnitude\_0\_1$ corresponds to the \textit{L2}-norm of the means of two packet size streams aggregated by traffic sent between a set of source and destination IP addresses with time window 5 (100 ms) and the feature $HpHp\_3\_covariance\_0\_1$ corresponds to the covariance between two packet size streams aggregated by traffic sent between a set of source and destination IP addresses with time window 3 (500 ms).

\begin{figure}[h]
\centering
\includegraphics[scale=0.4]{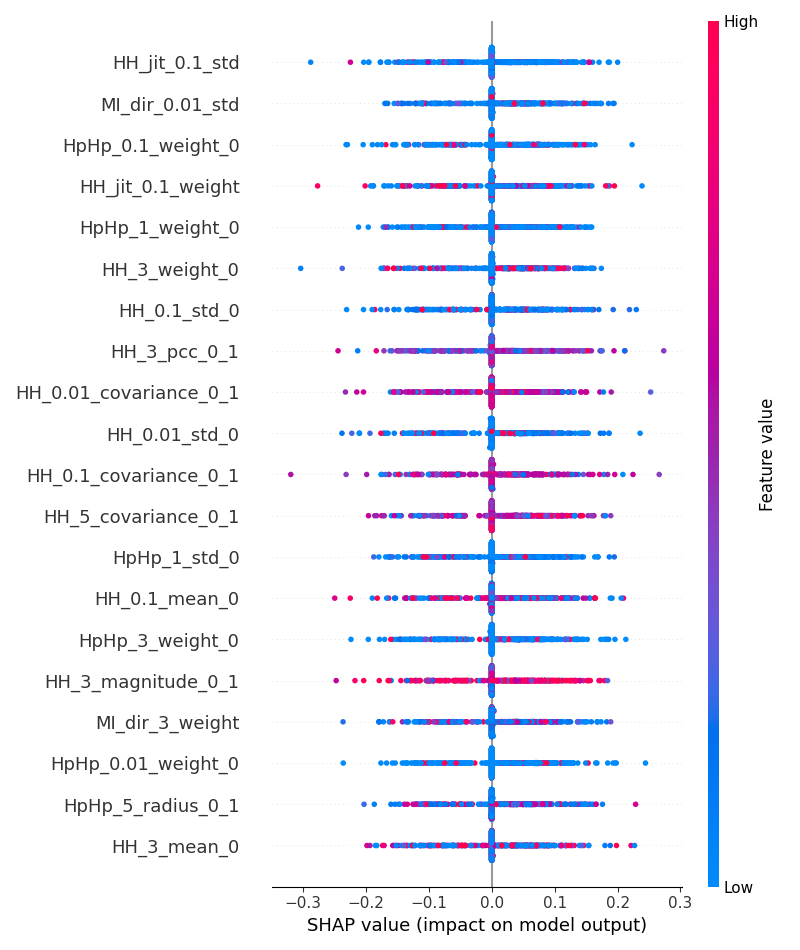}
\caption{SHAP beeswarm plot for HorusEye CICIDS-2017 model}
\label{horuseye-cicids-shap-beeswarm}
\end{figure}

\textbf{SHAP Analysis}: Using kernel SHAP analysis, Fig. \ref{horuseye-cicids-shap-beeswarm} shows a summary of how the top features in CICIDS-2017 dataset impact the HorusEye NIDS model’s output using a beeswarm plot for randomly selected 1000 data samples corresponding to the benign root node-leaf node path in the TRUSTEE DT (Fig. \ref{horuseye-cicids-dt}). Features such as $HH\_jit\_0.1\_std$, $MI\_dir\_0.01\_std$ and $HpHp\_0.1\_weight\_0$ have the most impact on the prediction value, in that order.

%Fig. \ref{horuseye-shap-beeswarm-trustee} shows the Shapley values for features shown in the major benign path for HorusEye's DT trained with the downsampled Mirai dataset using a SHAP’s beeswarm plot. Features such as $HpHp\_0.01\_pcc\_0\_1$, $HH\_1\_pcc\_0\_1$ and $HH\_3\_radius\_0\_1$ have the most impact on the prediction value, in that order. The feature $HH\_0.1\_weight\_0$ corresponds to weights aggregated by traffic sent between a set of source and destination IP addresses with time window 0.1 (10 seconds). It can be inferred from this figure that most Shapley values are negative, meaning that Shapley local interpretation agrees with TRUSTEE that, for the selected data samples, these six features’ values are pushing them toward being benign.

%\textbf{TRUSTEE vs SHAP Analysis}: Using the systematic approach for comparing TRUSTEE and SHAP explanations as explained in sub-section \ref{trustee-shap-rel}, the set of features used in the benign root node-leaf node path in the TRUSTEE DT is not a subset of the set of top contributing features computed by kernel SHAP summary analysis. Therefore, \textit{TRUSTEE AND SHAP explanations do not agree with each other at the subset level for HorusEye NIDS.}

\subsection{Case Study- ENIDrift $\times$ CICIDS-2017}
%\subsection{Case Study- ENIDrift}
\label{enidrift-case}
%Change the results to AE ensemble instead of PCA ensemble after journal review.

In this analysis, we used the CICIDS-2017-Wed (corresponding to the CICIDS data collected on July 5, 2017 (Wednesday)) dataset for training/testing ENIDrift NIDS.

\begin{figure}[h]
\centering
\includegraphics[scale=0.5]{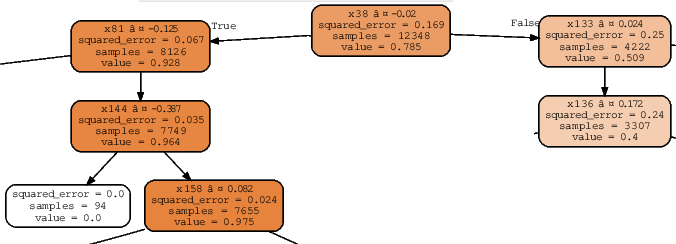}
\caption{Decision tree for ENIDrift CICIDS-2017 model with Top-k pruning (k=10). Only the top 4 layers are shown.}
\label{enidrift-cicids-dt}
\end{figure}

\textbf{TRUSTEE Analysis}: The results of TRUSTEE analysis for ENIDrift are shown in Table \ref{enidrift-cicids-trustee-table}. Using TRUSTEE with $30\%$ of the CICIDS-2017 dataset samples and no pruning results in the DT explanation which achieves 0.929 fidelity compared to ENIDrift. Using TRUSTEE with $30\%$ of the dataset samples and top-k Pruning method (setting $k = 10$) results in a DT explanation (shown in Fig. \ref{enidrift-cicids-dt}) which achieves 0.53 fidelity compared to ENIDrift. This is expected since pruning the branches of a DT reduces its fidelity. The most prominent features ENIDrift uses to determine anomaly are: $f38$, $f81$, $f158$.

\begin{table}[h]
	\centering
%	\vspace*{0.5cm}
    \begin{tabular}{ | l | l | l | l | }
    \hline
    \textbf{Sample size} & \thead{\textbf{Top-k pruning}\\ \textbf{used?}} & \textbf{DT size, depth, leaves} & \textbf{Fidelity} \\ \hline
    30\% & No & 253, 24, 127 & 0.929 \\ \hline
    30\% & Yes (k=10) & 25, 7, 13 & 0.53 \\ \hline
    \end{tabular}
    \caption{TRUSTEE analysis results for ENIDrift CICIDS-2017 model}
    \label{enidrift-cicids-trustee-table}
\end{table}

%\textbf{SHAP Analysis}: In Fig. \ref{enidrift-shap-beeswarm}, we show the local explanation for a malicious packet data point using the SHAP’s force plot for ENIDrift. The base value is $0.925$. The total positive contribution is greater than the negative contribution, and the final predicted value is greater than the base value. As a result, the class is predicted as malicious. A number of features have a positive and negative impacts on the prediction value. However, the kernel SHAP force plot does not show the feature names clearly, perhaps due to the large number of features having similar positive contributions and similar negative contributions to the prediction value. Again, we would like to point out that ENIDrift does not extract explicit features from incoming packets, instead it embeds them to vectors. So, the kernel SHAP force plot is in fact, showing the contributions of embedding vector components.

\begin{figure}[h]
\centering
\includegraphics[scale=0.4]{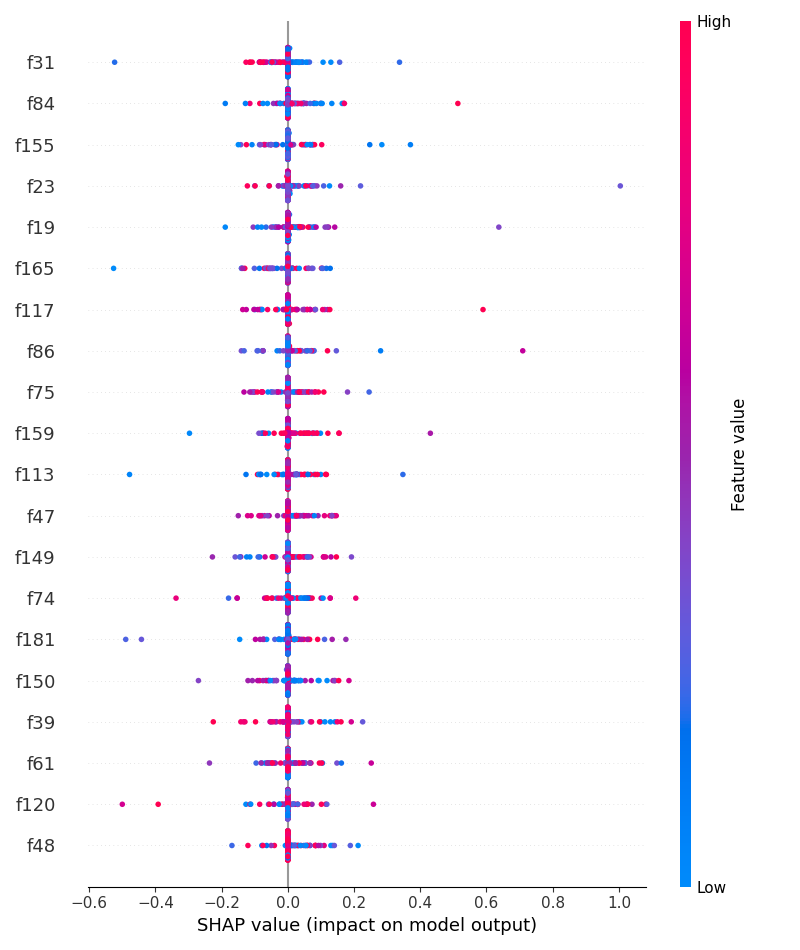}
\caption{SHAP beeswarm plot for ENIDrift CICIDS-2017 model}
\label{enidrift-cicids-shap-beeswarm}
\end{figure}

\textbf{SHAP Analysis}: Fig. \ref{enidrift-cicids-shap-beeswarm} shows a summary of how the top features in CICIDS-2017 dataset impact the ENIDrift NIDS model’s output using a beeswarm plot for randomly selected 300 data samples. Features (embedding vector components) such as $f31$, $f84$, $f155$ have the most impact on the prediction value, in that order.

%\textbf{TRUSTEE vs SHAP Analysis}: Using the systematic approach for comparing TRUSTEE and SHAP explanations as explained in sub-section \ref{trustee-shap-rel}, the set of features used in the benign root node-leaf node path in the TRUSTEE DT is not a subset of the set of top contributing features computed by kernel SHAP summary analysis (there are no common features between the two sets). Therefore, \textit{TRUSTEE AND SHAP explanations do not agree with each other at the subset level for ENIDrift NIDS.}

\subsection{Case Study- HELAD $\times$ CICIDS-2017}
%\subsection{Case Study- HELAD}
\label{helad-case}
In this analysis, we used a subset of the original CICIDS-2017 dataset consisting of 301,000 packets for this analysis as per the open source implementation \footnote{https://github.com/cdogemaru/CPIP} released by the authors. The first 200,000 packets were used for training and the rest of the packets were used for testing. Among the packets used for testing, the first 50,000 were benign while the rest were attack-related. The complete CICIDS-2017 dataset was not used since it increased the LSTM training time substantially. 

\begin{figure}[h]
\centering
\includegraphics[scale=0.35]{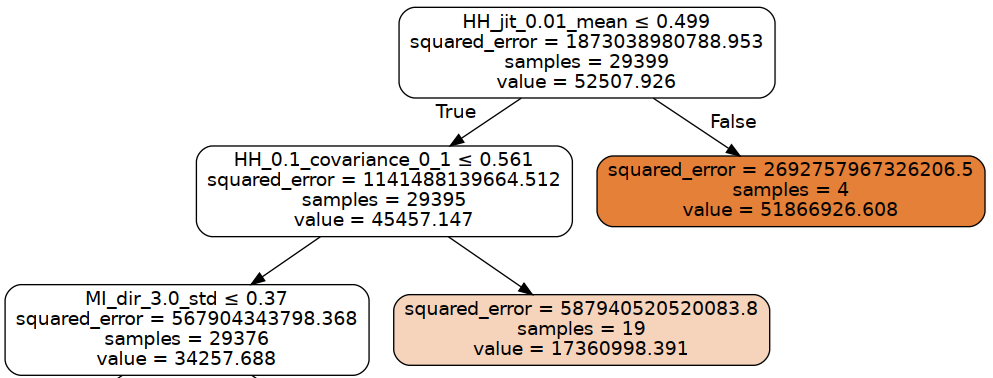}
\caption{Decision tree for HELAD CICIDS-2017 model with Top-k pruning (k=10). Only the top 3 layers are shown.}
\label{helad-cicids-dt}
\end{figure}

\textbf{TRUSTEE Analysis}: The results of TRUSTEE analysis for HELAD are shown in Table \ref{helad-cicids-trustee-table}. Using TRUSTEE with $30\%$ of the reduced CICIDS-2017 dataset samples and no pruning results in the DT explanation which achieves 0.0 fidelity compared to HELAD. Using TRUSTEE with $30\%$ of the dataset samples and top-k Pruning method (setting $k = 10$) results in a DT explanation (shown in Fig. \ref{helad-cicids-dt}) which achieves 0.0 fidelity compared to HELAD.

The top three prominent features HELAD's DT uses to determine an anomaly are:
\begin{enumerate}
    \item $HH\_jit\_0.01\_mean$- Packet jitter aggregated by traffic sent between a set of source and destination IP addresses with time window 0.01 (1 minute).
    \item $HH\_0.1\_radius\_0\_1$- \textit{L2}-norm of variances of two packet size streams aggregated by traffic sent between a set of source and destination IP addresses with time window 0.1 (10 seconds).
    \item $MI\_dir\_1.0\_std $- Standard deviation of packet sizes aggregated by source MAC and IP addresses, with time window 1.0 (1.5 seconds).
\end{enumerate}
Thus, \textit{HELAD's DT relies mainly on the jitter experienced and the sizes of packets exchanged between a pair of IP addresses per time frame to determine if an attack is underway.}

\begin{table}[h]
	\centering
%	\vspace*{0.5cm}
    \begin{tabular}{ | l | l | l | l | }
    \hline
    \textbf{Sample size} & \thead{\textbf{Top-k pruning}\\ \textbf{used?}} & \textbf{DT size, depth, leaves} & \textbf{Fidelity} \\ \hline
    30\% & No & 19607, 94, 9804 & 0.0 \\ \hline
    30\% & Yes (k=10) & 475, 77, 238 & 0.0 \\ \hline
    \end{tabular}
    \caption{TRUSTEE analysis results for HELAD NIDS}
    \label{helad-cicids-trustee-table}
\end{table}

\textbf{SHAP Analysis}: Running a kernel SHAP analysis of HELAD NIDS with the reduced CICIDS-2017 dataset resulted in issues. First, we ran the analysis with 200,000 packets but it threw a \textit{numpy} memory error since it was unable to allocate a huge amount of memory (313 GiB) for storing a ($2248 \times 20000000$) array with 64-bit float values. Next, we ran the analysis with 20,000 packets but the LSTM was taking too long to train. Hence, we decided to not proceed with the kernel SHAP analysis.

\end{document}